\documentclass[12pt]{iopart}

\usepackage{graphicx}

\newcommand{\Var}{\ensuremath{\mathrm{Var}}}

\begin{document}

%%%%%%%%%%%%%%% TITLE AREA %%%%%%%%%%%%%%% 

\title{Stochastic Models of Evolution in Genetics, Ecology and
Linguistics}

\author{R.\ A.\ Blythe} \address{SUPA, School of Physics, University
of Edinburgh, Mayfield Road, Edinburgh EH9 3JZ}

\author{A.\ J.\ McKane} \address{School of Physics and Astronomy,
University of Manchester, Manchester M13 9PL}

\begin{abstract}
We give a overview of stochastic models of evolution that have found
applications in genetics, ecology and linguistics for an audience of
nonspecialists, especially statistical physicists. In particular, we 
focus mostly on neutral models in which no intrinsic advantage is ascribed 
to a particular type of the variable unit, for example a gene, appearing in 
the theory. In many cases these models are exactly solvable and furthermore 
go some way to describing observed features of genetic, ecological and 
linguistic systems.
\end{abstract}

\maketitle

%%%%%%%%%%%%%%% MAIN TEXT %%%%%%%%%%%%%%% 

\section{Introduction}
\label{intro}

The distinguishing feature of statistical mechanics as a theory of
many-body physics is that it has probabilistic notions at its core.
At equilibrium, the principle of equal \textit{a priori} probabilities
is invoked as a means to identify the typical behaviour of a system as
a whole.  Out of equilibrium, the evolution of a system is usually
couched in terms of stochastic dynamics, as a glance at many articles
in this journal will confirm.  The procedure of stochastic modelling
is not, of course, restricted to physics.  In particular, stochastic
models have played a pivotal role in understanding the dynamics of
evolutionary systems and as such one sees many similarities in the
approaches and methods that have been used to those employed by
statistical physicists.  The most well-established discipline for
which this is the case is population genetics, i.e., the study of gene
frequencies among a population of reproducing individuals.  More
recently, models very similar to those that arise in the population
genetics arena have also been applied in ecology (where the
frequencies of particular species in an ecosystem are of interest) and
language change (where now the focus is on frequencies of linguistic
variants used by a community of speakers).  In this article, it is our
aim to introduce key concepts and results from these fields to
practitioners in statistical mechanics in order to further
cross-fertilisation between them.

Mostly, we will concentrate on the application to population genetics:
due to this discipline's age, most of the mathematical literature
describing the stochastic models of interest are couched in terms of
the language of population genetics.  We shall therefore begin by
introducing the relevant terminology.  In the simplified mathematical
models we shall be describing, a \emph{gene} is taken to mean a
variable unit that codes for a specific trait (such as eye colour),
and different variants of the gene (coding in this example for
different eye colours) are called \textit{alleles}.  Changes in allele
frequency can occur as a consequence of three processes.  First, there
is reproduction, in which offspring organisms acquire copies of
alleles from their parents.  Secondly, in the copying process, random
mutations can occur which may change one allele to another, or create
a completely new allele.  Finally, for whatever reason, organisms
carrying one allele may end up having more offspring than another:
this is selection.

The simplest stochastic model of an evolving population dates from the
1930's and was introduced independently by Ronald Fisher
\cite{Fisher30} and Sewall Wright \cite{Wright31}.  In this model, the
number of organisms is taken to be constant from one generation to the
next, and each instance of a gene in one generation is an exact copy
of one randomly chosen with replacement from the previous generation.
Such a process could be realised in a number of ways.  Let us consider
first of all \emph{haploid} organisms which carry only one instance of
each gene and (possibly) reproduce asexually.  A new generation could
then be formed by each of the organisms producing an infinite number
of \emph{gametes} (e.g., seeds) before dying; then, a random
fixed-size sample of these gametes survive to become offspring
organisms.  More often, geneticists focus on \emph{diploid} organisms
which carry two copies of each gene and reproduce sexually.  Now, each
organism produces two infinite sets of gametes, one for each allele
carried by the parent.  Then, after dying, the next generation is
formed by combining the requisite number of randomly-chosen gamete
pairs.  Either way, the mean number of offspring that any individual
has in the following generation is one, and so there is no selection:
this is a \emph{neutral} model.  Furthermore the fact that the allele
instances present in one generation are a random sample of those
present in the previous generation means that allele frequencies
fluctuate through sampling effects alone.  These fluctuations are
called \emph{genetic drift} and---in the absence of any
mutations---over time cause permanent extinction of one or more
alleles: The steady state of such a model is one in which only a
single allele remains: this allele is then said to have \emph{fixed}
in the population.

A haploid, or randomly-mating diploid population evolving in this way
is often referred to as an \emph{ideal} population.  In reality,
individuals do not mate at random, and there is often a preference, or
requirement, for mating to occur between or within different
\emph{classes} of individuals.  Most obviously, in sexually
reproducing organisms, females can mate only with males; other
factors, such as individuals' geographical locations and age (when one
has overlapping generations) may also lead to deviations from
non-ideal behaviour.  Nevertheless, in some cases it is found that the
predictions of the ideal model are relevant if one replaces the actual
size of the population with an effective size, which can be thought of
as a number of breeding individuals.  We will discuss this procedure
in more detail later in this article.

Neutral theories in population genetics (as well as in the ecological
and linguistic applications---see below) seem to have a far greater
realm of validity that might naively be expected.  The idea that many
genetic mutations are effectively neutral, that is do not
significantly affect the fitness of the carrier, was pioneered by
Motoo Kimura~\cite{Kimura68,Kimura83}.  Thus changes in their
frequency are due to chance, rather than selection. This is not to
deny that natural selection has an important role in the development
of morphological and behavioural characteristics, only that random
genetic drift has a more significant effect that had hitherto been
envisaged.  The relationship of the neutral theory to selection was
explored further with the introduction of the concept of
``near-neutrality''\cite{Ohta92,Ohta02}, in which the extent to which
genes are affected by drift and selection is a function of the
effective size of a breeding population.

In ecology the concept of neutrality is more recent, and is frequently
associated with Stephen Hubbell, who has developed and popularised the
concept during the last decade~\cite{Hubbell01}.  The idea of
neutrality in ecology can be explained by imagining a set of similar
species in a local community, trees in a forest, for example. When a
particular tree dies, it is replaced by an ``offspring'' of one of the
other trees in the forest---picked at random.  This is similar to
the Wright-Fisher model, except that now individuals (trees) are the
fundamental entities, not genes, and they come in different species,
rather than as different alleles. As in the genetic case, successive
random events will eventually lead to the extinction of one species by
chance, and then another, until eventually only one will remain. That
this does not happen is postulated to be due to new individuals
occasionally being introduced from outside the local community by
immigration. The distribution of species abundances calculated from
neutral theory gives a remarkably good fit to data~\cite{Condit02},
however the extreme simplicity of the model has made it very
controversial. Some ecologists see it as a good ``zeroth-order
approximation'' on which to build a more complete
description~\cite{Harte03}, while others believe some other kind of
starting point is required~\cite{McGill03,McGill06}.

The analogies to a model of language
change~\cite{Hull88,Hull01,Croft00,Croft02,Croft06} are a little more
abstract.  Here, there is a retained memory of usages of a particular
variant of a linguistic variable (such as a vowel sound or grammatical
structure); when a variant is reproduced, its retention in a speaker's
memory displaces a record of an earlier utterance.  Thus a speaker's
perception of the frequency with which a particular variant is
used---which in the quantitative version of the model which has been
developed~\cite{Baxter06}, is taken as the frequency with which that
same speaker produces that variant---also exhibits genetic-drift-like
fluctuations over time.  The imitation of linguistic variables
according to usage frequencies experienced by individual speakers
falls within a paradigm referred to as the \emph{usage-based model} in
linguistics \cite{Tomasello03} (this to it distinguish from theories
in which aspects of the grammar come ``pre-programmed'').
Investigations of the model of language change introduced in
\cite{Baxter06} have only just begun, but it is already clear that the
structure of the network of speakers, shaped by both geography and
social interactions, is important in determining the time needed for a
particular variant to become fixed.  In some cases, such as in new
dialect formation \cite{Gordon04,Trudgill04}, fixation of linguistic
variants has been seen to occur on the timescale of a human
generation: it is apparently possible that this may occur entirely
through random effects, without the need to invoke any selection
mechanism~\cite{Blythe07}.

In the next section of this article, we present a precise definition
of the model of neutral genetic drift in an ideal population, along
with the non-ideal generalisation of a subdivided populations.  We
also demonstrate concretely the relationship to the ecological and
linguistic models just described.  Then, in Sections~\ref{master} and
\ref{coalescent} we describe two complementary analytical approaches.
The first starts with a master equation and leads to a diffusion
approximation for allele frequencies; this type of approximation was
popular in the 1950s, largely through Kimura's efforts.  Although
implicitly used previously, the 1980s saw the formalisation of an
approach called the coalescent, in which the statistical properties of
the ancestry of a present-day population yield additional insights
into the population dynamics.  Finally, in Sections~\ref{nonideal} and
\ref{selection} we review some of the special features of non-ideal
and non-neutral evolution respectively, before making general remarks
in the conclusion, Section~\ref{conclusion}.

\section{Genetic drift: the Wright-Fisher and Moran models}
\label{ideal}

\subsection{The Wright-Fisher model}
\label{W-F}

\begin{figure}
\begin{center}
\includegraphics[width=0.66\linewidth]{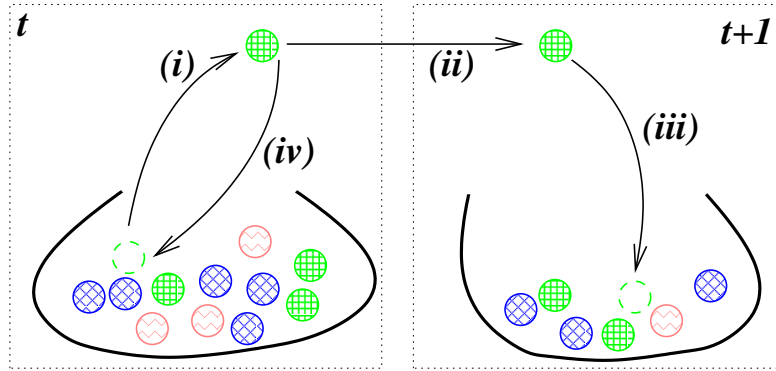}
\end{center}
\caption{\label{fig-beanbag} Wright-Fisher `beanbag' population
genetics.  The population in generation $t+1$ is constructed from
generation $t$ by (i) selecting a gene from the current generation at
random; (ii) copying this gene; (iii) placing the copy in the next
generation; (iv) returning the original to the parent population.
These steps are repeated until generation $t+1$ has the same sized
population as generation $t$.}
\end{figure}

The simplest, and most widely
known~\cite{Hartl89,Crow70,Ewens04,Maruyama77}, model of random
genetic drift is the Wright-Fisher
model~\cite{Fisher30,Wright31}. Suppose we have a population of
individuals in generation $t$ who mate randomly to produce the new
generation $t+1$. We focus on one particular gene which may be in one
of two states (alleles) which we call $A$ and $B$. The Wright-Fisher
model is based on the idea of the gene pool of the $t$ generation
individuals being sampled to give the genetic structure of the $t+1$
generation. In reality, as discussed in section \ref{intro}
individuals are diploid, but for convenience we assume that
individuals are haploid, that is, there is one-to-one correspondence
between individuals and genes. The process of constructing the new
generation from the current one is analogous to sampling coloured
balls---representing different alleles (types of gene)---from a bag or
urn. The steps in the process are illustrated in
Fig.~\ref{fig-beanbag}.

%%%%%%%%%%%%%%%%%%%%%%%%%%%%%%%%%%%%%%%%%%%%%%%%%%%%%%%%%%%%%%%%%%%%%%%%%%

%%%%%%%%%%%%%%%%%%%%%%%%%%%%%%%%%%%%%%%%%%%%%%%%%%%%%%%%%%%%%%%%%%%%%%%

Suppose that in generation $t$ there are $n'$ $A$ alleles and $(N-n')$
$B$ alleles. We wish to sample this gene pool $N$ times (with
replacement) in order to create the new generation $t+1$. The
probability that in $N$ trials of sampling the pool will get $n$ $A$
alleles is simply given by the binomial distribution:
\begin{equation}
{N \choose n}\,p_{1}^{n} p_{2}^{N-n}\,, 
\label{WF_binomial} 
\end{equation}
where $p_{1}(n')$ is the probability of picking an $A$ ($=n'/N $) and
$p_{2}(n')$ is the probability of picking a $B$ ($=(N-n')/N $). So the
probability that there is a transition from the original state
(characterised by $n'$ $A$ alleles) to the new state (characterised by
$n$ $A$ alleles) is given by
\begin{equation}
p_{n\,n'} = {N \choose n} \left( \frac{n'}{N} \right)^{n}
\left( 1 - \frac{n'}{N} \right)^{N-n}\,. 
\label{WF_trans_Probs}
\end{equation}
It should be noted that many mathematicians will write $p_{n'\,n}$
when they are referring to the transition probability from the state
$n'$ to the state $n$, whereas we follow the usual convention in
physics and denote this by $p_{n\,n'}$. This makes much of the matrix
algebra more natural---involving post-multiplication of vectors,
rather than pre-multiplication.

What we wish to calculate is the probability of finding $n$ $A$
alleles in generation $t$, given that there were $n_0$ initially. We
denote this probability by $P(n,t|n_{0},0)$, which we will frequently
abbreviate to $P(n,t)$, taking the initial condition as given. Since
the construction of the gene pool at generation $t+1$ only depends on
the state of the gene pool in generation $t$, and not on its state in
earlier generations, the process is Markov. Theoretical biologists,
like mathematicians, prefer to model such processes in terms of Markov
chains. Thus they define firstly, a probability vector
\begin{eqnarray*}
\underline{P} (t) =
\left(
\begin{array}{c}
P (1, t) \\
P (2, t) \\
P (3, t) \\
\vdots 
\end{array}
\right) 
\end{eqnarray*}
and secondly a transition matrix
\begin{eqnarray*}
{\cal P} =
\left(
\begin{array}{ccc}
p_{1 1} & p_{1 2} & \ldots \\
p_{2 1} & p_{2 2} & \ldots \\
\vdots  & \vdots & \ddots 
\end{array}
\right)\,. 
\end{eqnarray*}
The evolution of the system is then given by the matrix equation
\begin{equation}
\underline{P} (t+1) = {\cal P} \underline{P} (t) \ \ {\rm or} \ \ 
P(n, t+1) = \sum_{n'} p_{n\,n'} P(n', t)\,.  
\label{WF_chain}
\end{equation}
Note that the sum of the components of the probability vector must add up to 
1, and the sum of the entries in any one column of the transition matrix must 
add up to 1 (since $\sum_{n} P(n, t)=1$ and $\sum_{n} p_{n\,n'} = 1$). Matrices
whose column entries add up to 1 are known as \textit{stochastic matrices}, 
although the convention of writing the initial state as the index on the left
means that in the literature the statement is usually that the row entries add 
up to unity. It is now straightforward to formally express the state of the 
system in generation $t$ in terms of this initial state: 
\begin{equation}
\underline{P} (t) = {\cal P}\,\underline{P} (t-1) 
= {\cal P}\,{\cal P}\,\underline{P} (t-2) = \ldots = 
{\cal P}^{t} \underline{P} (0)\,.
\label{WF_formal}
\end{equation}
Clearly the large $t$ behaviour will be dominated by the largest (in magnitude)
eigenvalues of ${\cal P}$ and their corresponding eigenvectors. The eigenvalues
of the Wright-Fisher model are known~\cite{Feller51}, and are given by
\begin{equation}
\lambda_{r} = {N \choose r} \frac{r!}{N^{r}}\;,\quad r=0,\ldots,N\,.
\label{WF_eigenvalues} 
\end{equation}

Some simple deductions may be made from the Wright-Fisher model formulated
as a Markov chain. For instance, we may ask what the system looks like after
a large number of generations. If it tends to a stationary state, and if we 
call this stationary probability vector $\underline{P}^\ast$, it follows
that ${\cal P}\,\underline{P}^\ast = \underline{P}^\ast$. That is, 
$\underline{P}^\ast$ is a right eigenvector of ${\cal P}$ with eigenvalue 1.
Intuitively it seems clear that the final state is either all $A$ (the 
$A$ allele has become \textit{fixed}) or all $B$ (the $B$ allele is fixed).
This would correspond to a stationary state given by
\begin{eqnarray}
\underline{P}^\ast =
\left(
\begin{array}{c}
1-\Pi \\
0 \\
\vdots \\
0 \\
\Pi  
\end{array}
\right)\,.
\label{WF_stationary}
\end{eqnarray}
Here $\Pi$ is the probability that allele $A$ will eventually become fixed.
If we calculate ${\cal P}\,\underline{P}^\ast$ using the above expression, we 
find that it is indeed equal to $\underline{P}^\ast$, confirming that 
$\underline{P}^\ast$ is of this form.

Another simple deduction is that the expected number of $A$ alleles does not 
change from one generation to the next:
\begin{eqnarray}
\langle n(t+1) \rangle &=& \sum_{n} n P(n, t+1) = \sum_{n} n \sum_{n'} 
p_{n\,n'} P(n',t) \nonumber \\
&=& \sum_{n'} n' P(n',t) = \langle n(t) \rangle \,,   
\label{WF_mean}
\end{eqnarray}
since $\sum_{n} n p_{n\,n'} = N p_{1} = n'$. In particular, 
$\langle n(t) \rangle = \langle n(0) \rangle = n_{0}$. We can use this 
result to determine $\Pi$ in Eq.~(\ref{WF_stationary}). To do this we note from
Eq.~(\ref{WF_stationary}) that as $t \to \infty$ only two states are possible:
$n=N$ (with probability $\Pi$) and $n=0$ (with probability $1-\Pi$). Therefore
\begin{equation}
\lim_{t \to \infty} \langle n(t) \rangle = N \Pi + 0 (1-\Pi) \ \ \
\Rightarrow n_{0} = N \Pi\,.
\label{WF_find_Pi}
\end{equation}
This gives the explicit form for $\underline{P}^\ast$ and also tells us that 
the probability that eventual fixation of $A$ will occur is $n_{0}/N$.

\subsection{The Moran Model}
\label{Moran}

Statistical physicists are not used to formulating stochastic
processes in the way described above for the Wright-Fisher model; they
usually use continuous time master equations and define transitions
for a single stochastic step rather than for $N$ bundled
together~\cite{vanKampen92}. A variant of the Wright-Fisher model,
which is closer in spirit to stochastic processes considered by
statistical physicists, was introduced by Moran~\cite{Moran58}.  This
model does not have a fixed ``previous generation'' from which $N$
genes are sampled (with replacement) to form the next
generation. Instead, genes are chosen to die one at a time, and
replaced by new ones which are $A$ with probability $p_1$ and $B$ with
probability $p_2$.  These probabilities are calculated from the state
of the system prior to the death of the chosen individual: $p_{1}(n) =
n/N$ and $p_{2}(n)= (N-n)/N$, where $n$ is the number of $A$ alleles
just before the combined birth-death event.

The Moran model does not have non-overlapping generations, as in the 
Wright-Fisher model: after $N$ sampling events in the Moran model, each gene 
need not have been replaced, and several might not have survived very long at 
all. It is therefore said to have overlapping generations. Due to the 
impossibility of unambiguously identifying a ``generation'' in this case, a 
generation in the Moran model is frequently defined as the period over which a 
single sampling event takes place. This is a confusing use of the word 
``generation'', but at least it is well defined. So, in one generation the 
population of $A$ may go up by one (if a $B$ is chosen to die, and an $A$ is 
chosen to be born) or down by one (if an $A$ is chosen to die, and a $B$ is 
chosen to be born). The number of $A$ alleles will remain unchanged if both an 
$A$ is chosen to die and to be born, and also if a $B$ is chosen to die and to 
be born. This gives the transition matrix to be
\begin{equation}
\label{M_p}
p_{n'\,n} = \left\{ \begin{array}{ll} 
\left( 1 - \frac{n}{N} \right) p_{1}(n)\,, & 
\mbox{\ if $n'=n+1$} \\ 
\left( \frac{n}{N} \right) p_{1}(n) + \left( 1 - \frac{n}{N} \right) 
p_{2}(n)\,, & \mbox{\ if $n'=n$} \\
\left( \frac{n}{N} \right) p_{2}(n)\,, & 
\mbox{\ if $n'=n-1$} \\
0\,, & \mbox{\ otherwise\,,}
\end{array} \right.
\end{equation}
from which it follows that $\sum_{n'} p_{n'\,n} = 1$. 

The explicit expressions for the non-zero elements of the transition matrix 
of the Moran model are therefore
\begin{eqnarray}
p_{n+1\,n} &=& \left( 1 - \frac{n}{N} \right) \left( \frac{n}{N} \right)\,,
\nonumber \\
p_{n\,n} &=& \left( \frac{n}{N} \right)^{2} + 
\left( 1 - \frac{n}{N} \right)^{2}\,, \nonumber \\
p_{n-1\,n} &=& \left( \frac{n}{N} \right) \left( 1 - \frac{n}{N} \right)\,.
\label{M_p_nonzero}
\end{eqnarray}

The transition matrix (\ref{M_p}) for the Moran model is much simpler than 
for the Wright-Fisher model, being tri-diagonal. Therefore it is not surprising
that the eigenvalues of the transition matrix (\ref{M_p}) can, as for the
Wright-Fisher model, found exactly~\cite{Moran58}:
\begin{equation}
\lambda_{r} = 1 - \frac{r(r-1)}{N^2}\;,\quad r=0,\ldots,N \,,
\label{M_eigenvalues} 
\end{equation}
but, unlike the Wright-Fisher model, the corresponding eigenvectors are also 
known~\cite{Moran62}. The two largest eigenvalues are both unity; they have 
left eigenvectors $(1,1,\ldots,1)$ and $(0,1,\ldots,N)$ and corresponding 
right eigenvectors $(1,0,0,\ldots,0)^{\rm T}$ and $(0,0,\ldots,0,1)^{\rm T}$. 
Clearly the last two column vectors correspond to the loss and fixation of the 
$A$ allele. The third largest eigenvalue, $\lambda_{2} = 1 - (2/N^{2})$ has 
the left eigenvector $(0,N-1,2(N-2),\ldots,n(N-n),\ldots,0)$ and right 
eigenvector 
$(-(1/2)(N-1),1,1,\ldots,1,-(1/2)(N-1))^{\rm T}$~\cite{Moran58,Moran62}. This
column vector governs the  behaviour, after a large number of generations, of 
the non-fixed states.  In particular, we have, once $t$ is sufficiently large,
\begin{equation}
\underline{P}(t) \approx \left( \begin{array}{c} 1 - \frac{n_0}{N} \\ 0 \\
\vdots \\ 0 \\ \frac{n_0}{N} \end{array} \right) + \frac{6 n_0
  (N-n_0)}{N(N^2-1)} \left( \begin{array}{c} - \frac{(N-1)}{2} \\ 1
  \\ \vdots \\ 1 \\ -\frac{(N-1)}{2} \end{array} \right) \left( 1 -
\frac{2}{N^2} \right)^t 
\end{equation}
where $n_0$ is the initial number of $A$ alleles.  We see that the
unfixed states are equally populated except at the boundaries, in
agreement with graphical representations of the pdf displayed in,
e.g.,~\cite{Wright31,Crow70}, which show it to be essentially flat away from
the boundaries after many generations.

\subsection{Mutations in the Wright-Fisher and Moran Models}
\label{WF_M_mut}
So far we have been considering situations where the change in composition 
of the population is caused by pure genetic drift---randomness 
with no underlying deterministic behaviour. We now include the effects of
mutation: an $A$ allele may mutate with a probability of $u$ to 
a $B$ allele, and a $B$ allele may mutate with a probability of $v$ to an $A$ 
allele. These are probabilities per generation, so once again these quantities 
in the Wright-Fisher model will differ from those in the Moran model by a 
factor of $N$. 

To include mutation, one chooses an allele from the current population to 
die, and replaces it with one of a type chosen also with a probability 
proportional to the current population sizes. This is as in the original 
models. The mutational stage is inserted at the end of the process, so the 
chosen successor is allowed to mutate before being introduced into the system 
to define the new population. In other words, when we pick an allele to be 
the ``parent'' of a ``child'' in the next generation, the offspring can mutate 
with the probabilities $u$ or $v$. For instance, if an $A$ allele is chosen, 
there is a probability of $(1-u)$ that the replacement allele is also an $A$ 
and of $u$ that it is a $B$. So, with mutation added,
\begin{equation}
p_{1}(n) = \left( 1-u \right) \left( \frac{n}{N} \right) 
+ v \left( 1 - \frac{n}{N} \right)\,,
\label{Mut_p1}
\end{equation}
and
\begin{equation}
p_{2}(n) = u \left( \frac{n}{N} \right) + \left( 1-v \right) 
\left( 1 - \frac{n}{N} \right)\,.
\label{Mut_p2}
\end{equation}
The transition matrices for the Wright-Fisher and Moran models are again found 
from Eqs.~(\ref{WF_binomial}) and (\ref{M_p}), but using $p_1$ and $p_2$ given 
by Eqs.~(\ref{Mut_p1}) and (\ref{Mut_p2}). So for instance, the non-zero 
elements of the transition matrix for the Moran model are
\begin{eqnarray}
p_{n+1\,n} &=& (1-u) \left( 1 - \frac{n}{N} \right) \left( \frac{n}{N} \right)
+ v \left(1 - \frac{n}{N} \right)^{2}\,, \nonumber \\
p_{n\,n} &=& (1-u) \left( \frac{n}{N} \right)^{2} + (1-v)
\left( 1 - \frac{n}{N} \right)^{2} + (u+v)\left( \frac{n}{N} \right)
\left( 1 - \frac{n}{N} \right)\,, \nonumber \\
p_{n-1\,n} &=& (1-v) \left( \frac{n}{N} \right) \left( 1 - \frac{n}{N} \right)
+ u \left( \frac{n}{N} \right)^{2}\,.
\label{M_p_nonzero_Mut}
\end{eqnarray}

Allowing for the possibility of mutation means that the expected number of 
$A$ alleles is no longer conserved. For the Wright-Fisher model this can be
calculated as in Eq.~(\ref{WF_mean}), but now 
\begin{equation}
\sum_{n} n p_{n\,n'} = N p_{1} = (1-u) n' + v (N-n')\,,
\label{WF_Mut_mean}
\end{equation}
and so
\begin{equation}
\langle n(t+1) \rangle = (1-u) \langle n(t) \rangle  + v \left( N 
-\langle n(t) \rangle \right)\,.
\label{WF_Mut_deter}
\end{equation}
For the Moran model we calculate $\sum_{n} n\,p_{n\,n'}$ directly from 
Eq.~(\ref{M_p_nonzero_Mut}) and find that
\begin{equation}
\langle n(t+1) \rangle = \langle n(t) \rangle + 
v \left( 1 - \frac{\langle n(t) \rangle}{N} \right) 
- u \left( \frac{\langle n(t) \rangle}{N} \right)\,,
\label{M_Mut_deter}
\end{equation}
which is of the same form as Eq.~(\ref{WF_Mut_deter}), apart from the factors
of $N$ dividing the mutation probabilities. Eqs.~(\ref{WF_Mut_deter}) and
(\ref{M_Mut_deter}) give the number of $A$ alleles in an infinite population, 
where no fluctuations are present. Written in terms of allele frequencies,
$a_{t} = \langle n(t) \rangle/N$ and $b_{t} = 1 - a_{t}$, 
Eq.~(\ref{WF_Mut_deter}) is nothing else but the deterministic equation for 
mutational change, $a_{t+1} = (1-u) a_{t} + v b_{t}$, found in many textbooks 
on population genetics~\cite{Hartl89,Crow70,Ewens04}. 

This difference equation for $a_t$ may be easily solved by writing 
$a_{t} = v/(u+v) + c_{t}$, so that $c_{t+1} = (1-u-v) c_{t}$. The equation 
for $c_t$ can be solved iteratively to give $c_{t} = (1-u-v)^{t} c_{0}$, which
yields
\begin{eqnarray}
a_{t} &=& \frac{v}{u+v} + \left( 1 -u-v \right)^{t} \left[ a_{0} - 
\frac{v}{u+v} \right] \nonumber \\
&=& \frac{v}{u+v} \left[ 1 - \left( 1 - u - v \right)^{t} \right] + 
a_{0}\left( 1-u-v \right)^{t}\,.
\label{soln_a_t}
\end{eqnarray}
Since $0 < 1-u-v < 1$, $a_{t} \to v/(u+v)$ as $t \to \infty$.

The addition of mutations also means that neither the $A$ nor the $B$ allele 
can become fixed---there is always the possibility of an allele being created
by mutation, even if there are currently none present in the population. Thus
the behaviour of both the Wright-Fisher and Moran models with mutation, after 
a large number of generations, is more complicated to obtain. Nevertheless, 
the eigenvalues of the transition matrices are known. For the Wright-Fisher 
model~\cite{Feller51}:
\begin{equation}
\lambda_{r} = \left( 1 - u - v \right)^{r} {N \choose r}
\frac{r!}{N^{r}} \;,\quad r=0,\ldots,N \,,
\label{WF_Mut_eigenvalues} 
\end{equation}
and for the Moran model~\cite{Ewens04}
\begin{equation}
\lambda_{r} = 1 - (u+v)\frac{r}{N} - \left( 1 - u - v \right) \frac{r(r-1)}
{N^2}\;, \quad r=0,\ldots,N \,.
\label{M_Mut_eigenvalues} 
\end{equation}
The Moran model with mutation can be solved
analytically~\cite{Karlin62,Karlin64}. In addition, approximate 
methods exist, such as the diffusion approximation discussed in 
Section \ref{master}, which describe these models well. Finally, all of
the models can be generalised to the cases with two sexes and diploid
individuals~\cite{Moran62}. These models are more complicated, but do
not in our view make the underlying ideas or the connection between
different approaches any clearer.  For all these developments we refer
the reader to the literature on the
subject~\cite{Moran62,Crow70,Maruyama77,Ewens04}; meanwhile, in
Section~\ref{nonideal} we shall examine how sexual reproduction and
diploidy are handled in models that are more easily studied analytically.

\subsection{Migration between Islands}
\label{WF_M_mig}

In the models of genetic drift considered so far, new individuals 
were produced from their ``parents'' randomly. If the organisms had been 
diploid, we would have said that the mating was random. This randomness 
assumption may model some situations, for example, if sperm is released into a 
pond or lake by a group of male fish in order to fertilise eggs laid by females
nearby. However, there are many instances where mating is not random. One of 
the most interesting is when groups of individuals are geographically 
separated. Then, if there is little migration between the populations, so that 
they are effectively isolated, stochastic effects mean that they will 
develop in different ways. For instance, some alleles may be lost in some of 
the populations, but not in others. 

Models which describe the population genetics of subdivided
populations which are partly isolated from each other are called
``island models'' and were first investigated by
Wright~\cite{Wright31}. The island populations themselves are referred
to as \textit{demes}. In early models, migration between demes was
essentially independent of the distance between the islands which
contained them, that is, each deme interchanged genes equally with
every other deme. More elaborate models soon followed, such as
stepping stone models, where the rate of migration between demes is a
function of the distance between the
islands~\cite{Kimura64,Malecot50,Malecot62}. These models can be
generalised by defining a migration matrix, $g_{ij}$, which specifies
the probability that, if an individual in deme $j$ has had an offspring,
the later immediately migrates to deme $i$ (so that it has arrived at
the start of the next generation).

The inclusion of migration into the models so far considered in this
section is in principle straightforward, but in some cases the matrix
of transition probabilities becomes cumbersome to write down. There
are also potentially several different models which can be constructed
depending on the order in which the various processes of birth,
mutation and migration take place, how many migrants are permitted
into and out of each deme, and so on. The introduction of migration
into Moran-type models is, as expected, easier to formulate and seems
more natural to statistical physicists. In the usual spirit of this
approach, migrations will be considered as single events, for
instance, a migration of a gene from deme $j$ to deme $i$. This is in
contrast to many other approaches to stochastic models of migration,
including one introduced by Moran~\cite{Moran59}, where a fixed number
of individuals are chosen to migrate from one deme to another.  We
shall consider the class of models for which migration can be
characterised through the matrix $g_{ij}$ defined above: this class
includes the simple cases considered by early workers, such as uniform
migration rate between all demes or stepping stone models.  The number
of demes will be taken to be $L$ and we will assume that they will
each consist of $N$ genes. For simplicity we will assume that the
mutation rates will be the same on each island. In other words, each
island will be an exact copy of the systems considered previously in
this section, but coupled together by migration processes.

To illustrate the basic idea, suppose that there are only two islands,
not $L$ as assumed above. Since each island contains exactly $N$
genes, only two integers are required to specify the state of the
system: $n_1$ and $n_2$, where $n_i$ is the number of $A$ alleles in
deme $i$, where $i=1,2$. We begin with the basic Moran model of
section \ref{Moran} where only the birth/death process is
considered. When migration is added, a ``death'' is not necessarily
followed by a ``birth''---it can be followed by a migration event. The
algorithm we use is as follows.  First, we randomly pick a ``parent'',
or gene to be copied, from deme $i$ with probability $f_i$.  Suppose
this parent is on island $2$.  Then with probability $g_{1 2}$ a
migration event takes place, which entails a death on island 1, and
the copied gene from island 2 occupying the space that was left
behind. With probability $g_{22} = 1-g_{1 2}$ we assume that no
migration event takes place and a death takes place on island 2, with
the copied gene from island 2 replacing it.  Note that the total
probability per generation that an individual migrates from deme $i$
to deme $j$ per generation is $g_{ji} f_i$.

Since $N$ is fixed for each deme, if $B$ increases by $1$ on island
$i$, this is equivalent to $A$ decreasing by $1$ on island $i$. So the
four independent transition probabilities are given by
\begin{eqnarray}
p_{(n_{1}+1,n_2)\,(n_{1},n_2)} &=& f_1 \left( 1 - g_{2 1} \right) 
\left( 1 - \frac{n_{1}}{N} \right) \frac{n_1}{N} + f_2 g_{1 2} 
\left( 1 - \frac{n_1}{N} \right) \frac{n_2}{N}\,, \nonumber \\
p_{(n_{1}-1,n_2)\,(n_{1},n_2)} &=& f_1 \left( 1 - g_{2 1} \right) 
\left( 1 - \frac{n_1}{N} \right) \frac{n_1}{N} + f_2 g_{1 2} 
\frac{n_1}{N} \left( 1 - \frac{n_2}{N} \right)\, \nonumber \\
p_{(n_{1},n_{2}+1)\,(n_{1},n_2)} &=& f_2 \left( 1 - g_{1 2} \right) 
\left( 1 - \frac{n_2}{N} \right) \frac{n_2}{N} + f_1 g_{2 1} 
\left( 1 - \frac{n_2}{N} \right) \frac{n_1}{N} \nonumber \\
p_{(n_{1},n_{2}-1)\,(n_{1},n_2)} &=& f_2 \left( 1 - g_{1 2} \right) 
\left( 1 - \frac{n_2}{N} \right) \frac{n_2}{N} + f_1 g_{2 1} 
\frac{n_2}{N} \left( 1 - \frac{n_1}{N} \right)\,,
\label{M_mig_probs}
\end{eqnarray}
with
\begin{eqnarray}
p_{(n_{1},n_{2})\,(n_{1},n_{2})} &=& 1 - p_{(n_{1}+1,n_2)\,(n_{1},n_2)} -
p_{(n_{1}-1,n_2)\,(n_{1},n_2)} \nonumber \\
&-& p_{(n_{1},n_{2}+1)\,(n_{1},n_2)} - p_{(n_{1},n_{2}-1)\,(n_{1},n_2)}\,. 
\label{M_mig_probs_last}
\end{eqnarray}

We can obtain the deterministic equation, valid in the limit $N \to \infty$,
describing the effect of migration on the number of $A$ alleles on one island,
by using the same method as was used to find Eq.~(\ref{M_Mut_deter}), 
describing the effects of mutation. In this case we calculate
$\sum_{n_{1},n_{2}} n_{1}\,p_{(n_{1},n_{2})\,(n_{1}',n_{2}')}$ from 
Eqs.~(\ref{M_mig_probs}) and (\ref{M_mig_probs_last}) to find that
\begin{equation}
\langle n_{1}(t+1) \rangle = \langle n_{1}(t) \rangle + 
\frac{f_2 g_{12}}{N} \left[ \langle n_{2}(t)\rangle  -  
\langle n_{1}(t) \rangle \right] \,,
\label{M_Mig_deter}
\end{equation}
which is of the same form as Eq.~(\ref{M_Mut_deter}) for mutations.
The corresponding expression for the expected number of $A$ alleles on
island $2$, $\langle n_{2}(t+1) \rangle$ is obtained by exchanging all
$1$ and $2$ indices.  Using these expressions one can show that the
weighted mean
\begin{equation}
\label{barn}
\bar{n}(t) \equiv f_1 g_{21} \langle n_1(t) \rangle + f_2 g_{12}
\langle n_2(t) \rangle
\end{equation}
is conserved, i.e., $\bar{n}(t+1)=\bar{n}(t)$, whilst the difference
\begin{equation}
\delta n(t) \equiv \langle n_1(t) \rangle - \langle n_2(t) \rangle
\end{equation}
decays geometrically since
\begin{equation}
\delta n(t+1) = \left[ 1 - \frac{f_1 g_{21} + f_2 g_{12}}{N} \right]
\delta n (t) \;.
\end{equation}
In other words, migration results in a convergence of the mean allele
frequencies on the two islands to a common value which in turn
eventually reaches a stationary value $n^\ast$ that can be worked
out from equation (\ref{barn}) once the initial values $n_1(0)$ and
$n_2(0)$ are known:
\begin{equation}
n^\ast = \frac{f_1 g_{21} n_1(0) + f_2 g_{12} n_2(0)}{f_1 g_{21} +
f_2 g_{12}} \;.
\end{equation}

\subsection{Mapping between population genetic, ecological and linguistic 
models}
\label{PG_E_L}
In section \ref{intro} we have already discussed the analogies that can be 
drawn between models of population genetics, individual based models (IBMs)
in ecology and models of the evolution of language. The mathematical formalism
we have introduced in the current section allows us to make these analogies 
more concrete, and to define mappings between models in these three distinct 
areas.

The genetic models which we have described have been of a gene at a single 
position (locus) on a chromosome, involved only one chromosome, and has not
included sex: new genes were ``born'' from a single parent. In this very
simplified genetics, we can directly identify a gene with an individual and 
vice-versa. In this case different alleles can be thought of as different
species. Therefore in the context in which we have been working, the mappings 
[genes $\leftrightarrow$ individuals] and [alleles $\leftrightarrow$ species] 
is very natural. The idea of an island is common to both population genetics 
and population ecology, where it is called a patch, and we have the further 
correspondence [deme $\leftrightarrow$ local community]. 

To extend these mappings to the evolutionary model of language 
change~\cite{Hull88,Hull01,Croft02,Croft06} described briefly in 
section \ref{intro}, we begin by noting that the central idea of this approach 
is to make an analogy between 
``different ways of saying the same thing''---called \textit{variants} of a 
word or set of words, and alleles. The specific ``word or set of words'' under
consideration is called the \textit{lingueme}. Thus the first analogy is
[variants $\leftrightarrow$ alleles]. In the model~\cite{Baxter06}, each 
speaker's utterances may contain different variants of a lingume. In the case 
of two variants, the speaker's grammar can be modelled as being made up of $N$ 
\textit{tokens}, of which $n$ are of one variant and $N-n$ of the other. An 
utterance by speaker $i$ will then produce tokens which modifies his own 
grammar (death and birth of tokens belonging to speaker $i$) and which may
modify the grammar of speaker $j$ (migration of tokens from speaker $i$ to 
speaker $j$). This gives the further analogies [tokens $\leftrightarrow$ genes]
and [speakers $\leftrightarrow$ islands]. Table \ref{mappings} summarises the
mappings between these different areas.

%%%%%%%%%%%%%%%%%%%%%%%%%%%%%%%%%%%%%%%%%%%%%%%%%%%%%%%%%%%%%%%%%%%%%%%%%
%%%%%%%%%%%%%%%%%%%%%%%%%%%%%%%%%%%%%%%%%%%%%%%%%%%%%%%%%%%%%%%%%%%%%%%%%

\begin{table}
\begin{center}\begin{tabular}{|c|c|c|}
\hline 
Population genetics&
Population ecology&
Language evolution\tabularnewline
\hline
\hline 
gene&
individual&
token\tabularnewline
\hline 
allele&
species&
variant\tabularnewline
\hline 
island&
patch&
speaker\tabularnewline
\hline 
deme&
local community&
grammar\tabularnewline
\hline
population&
metacommunity&
speech community\tabularnewline
\hline
mutation&
speciation&
innovation\tabularnewline
\hline
migration&
immigration&
conversation \tabularnewline
\hline
\end{tabular}\end{center}

\caption{The correspondences between constituents, concepts and processes in
population genetics, population ecology and language 
evolution.\label{mappings}}
\end{table}

%%%%%%%%%%%%%%%%%%%%%%%%%%%%%%%%%%%%%%%%%%%%%%%%%%%%%%%%%%%%%%%%%%%%%%%%%%%
%%%%%%%%%%%%%%%%%%%%%%%%%%%%%%%%%%%%%%%%%%%%%%%%%%%%%%%%%%%%%%%%%%%%%%%%%%%

While we have so far assumed that there are only two alleles, $A$ and $B$, 
in the case of population genetics, the analogies with population ecology 
and evolutionary linguistics make it clear that we will frequently be 
interested in having more than two species or variants in our models. 
Therefore in the next section will generalise the models so that there are 
$M$ possible types of gene/individual/token.  

The interpretation of ``genetic drift'', mutation and migration in
population ecology and evolving linguistics is interesting. The
species being considered in the analogy are not to be thought of as
predators and prey, but rather as species of a similar type
(``trophically similar species'') competing for a common resource,
such as space or light. The fact that when a death of an individual
occurs, it is immediately replaced by another (either through a birth
or by immigration) means that there is intense competition; as soon as
a ``vacancy'' occurs, it is immediately filled. This coupling of the
death rate to the birth/immigration rate so that the overall
population remains fixed, is called the ``zero-sum rule'' in biology.
Models where trophically similar species (for example, forests of
trees) change their population sizes through purely random dynamics
analogous to pure genetic drift, are called neutral
models~\cite{Hubbell01}. They are, at present, quite
controversial~\cite{Alonso06}, mainly because measurable ``physical''
quantities such as the species abundance distribution arise as a
result of random effects. In the context of neutral theories a
mutation corresponds to speciation, where a single individual of a
species changes into an individual of a different species. This is, of
course, a very coarse-grained description of a very complex
process. Immigration has the same interpretation as migration in the
genetic case.

In the linguistics application, we think of different linguistic
variants competing for use by members of the speech community.  Here,
the analogue of ``trophically similar species'' would be functionally
equivalent variants, or ``two ways of saying the same thing''.  For
true neutrality, one would further require that no speakers ascribe
any social value (``prestige'') to a variant, which might cause them
to reproduce it more (or, indeed, less) often they have heard it used
thus far.  There are many ways in which new variants might be created,
a process referred to as \emph{innovation} \cite{Croft00}.  Basically,
most of these arise from speakers needing to communicate something
that they have not communicated before; other forces behind innovation
include articulatory constraints which may involve speakers producing
one variant, even if they aim to produce another.  This latter case,
in particular, could be modelled by mutation rates between variants,
as we have described for the genetics application.  Migration
corresponds to the process whereby tokens from the grammar of one
speaker are incorporated into the grammar of another speaker with whom
the first is conversing.  Two factors contribute to the overall rate
at which tokens migrate from one grammar to another: first, the is the
frequency with which two speakers converse; secondly, speakers may
give higher or lower weights to the utterances of other speakers
based, for example, on the listener's perception of how similar the
interlocutor is to themselves, or whether they wish to acquire social
status by imitating variants associated with a prestigious group.
Such sociolinguistic effects are discussed in more detail in
\cite{Croft00}.  Meanwhile, it has been proposed \cite{Trudgill04}
that when dialects are brought into contact, as happened for example
when people from Britain and Ireland migrated to New Zealand, a truly
neutral theory (i.e., one in which these social factors are claimed to
be unimportant) should describe observed effects, such as convergence
to a relatively homogeneous dialect among a large community of
speakers on a timescale of order of a human lifetime.  Whether a
genetic drift-like process actually allows for such behaviour is
another matter of debate and is under current investigation.  

In this section we have introduced the simplest stochastic models of
population genetics and brought out analogies with similar models in
population ecology and evolutionary linguistics. In the next section
we will write these models in a form more familiar to statistical
physicists, generalise them, and also show that when the number of
genes/individuals/tokens for an island/patch/speaker is large, they
may be replaced by models based on Fokker-Planck equations, rather
than on master equations.

\section{Master and Fokker-Planck equations}
\label{master}

As we have seen in Section \ref{ideal}, the Wright-Fisher model is
quite difficult to analyse, especially when mutation and migration are
allowed for, and so very early on workers resorted to the
\textit{diffusion approximation}~\cite{Fisher22,Wright45,Kimura55a}.
This is a description of the process, valid when $N$ is large, where
(a) the (rational) allele frequencies $n/N$ are replaced by real
numbers $x$, $0 \leq x \leq 1$, and (b) the generational label, $t$,
is rescaled by a factor of $1/N$. Specifically, $t$ now measures units
of $N$ generations, so that one generation becomes only $1/N$ units of
$t$. The governing equation is now a Fokker-Planck equation for
$P(x,\tau'|x_{0},0)$ which for pure genetic drift has the form
\begin{equation}
\frac{\partial P}{\partial \tau'} = \frac{1}{2} 
\frac{\partial^{2}}{\partial x^2} \left[ x(1-x) P \right]\,,
\label{diffusion}
\end{equation}
where $\tau'=t/N$. In essence, this is a mesoscopic description, whereas the 
Wright-Fisher equation was a microscopic, or gene-based description. It is 
possible to add terms to Eq.~(\ref{diffusion}) which account for mutations and 
migrations.

In the last section we also described how the Moran model, while
similar to the Wright-Fisher model, was in many ways easier to write
down and generalise.  Therefore in this section we will formulate the
basic stochastic processes that interest us in terms of Moran-type
models, rather than Wright-Fisher models.  An advantage is that
Moran-type models may be formulated as master equations, a step that
is not typically followed in the mathematical biology literature.  The
advantage of formulating a stochastic process in this way, is that
there there then exist systematic methods to approximate the master
equation for large $N$ as a Fokker-Planck
equation~\cite{vanKampen92}. It will turn out that the Fokker-Planck
equation corresponding to the large $N$ limit of the Moran model has
the same form as Eq.~(\ref{diffusion}), that is, has the same limit as
the Wright-Fisher model.

We will begin by discussing the reformulation of the Markov chain
based description of population genetics given in Section \ref{ideal}
to the master equation description which we will use in the rest of
the section.

\subsection{Master equations and population genetics}
\label{continuous_time}

We begin by rewriting the standard dynamical relation for a Markov
chain, $P(n, t+1) = \sum_{n'} p_{n\,n'} P(n', t)$ as
\begin{equation}
P(n, t+1) - P(n, t) = \sum_{n'} p_{n\,n'} P(n', t) 
- \sum_{n'} p_{n'\,n} P(n, t)\,,
\label{markov}
\end{equation}
using $\sum_{n'} p_{n'\,n} = 1$. Now the terms in the sums with $n'=n$
may be omitted---since they cancel between the two terms on the
right-hand side---and so only the terms involving an \textit{actual
transition}, that is, terms with $n' \neq n$, appear in the equation.

Up until this point, we have been careful not to refer to $t$ as a
``time''---it has simply been an integer labelling
generations. However now we interpret $t$ as time divided into
intervals $\Delta t$ taken to be sufficiently small that at most one
sampling event occurs during it.  Dividing Eq.~(\ref{markov}) by
$\Delta t$, and omitting the terms with $n'=n$, yields
\begin{equation}
\frac{P(n, t+\Delta t) - P(n, t)}{\Delta t} = \sum_{n' \neq n} 
\left( \frac{p_{n\,n'}}{\Delta t} \right) P(n', t) - 
\sum_{n' \neq n} \left( \frac{p_{n'\,n}}{\Delta t} \right) P(n, t)\,.
\label{markov_master}
\end{equation}
We now adopt a different view of the stochastic process. Instead of
assuming that one exactly one sampling event happens per generation
(including $n \to n$, where no transition actually occurs), we now
have sampling events occur at unit rate so that \emph{on average} one
event takes place per generation.  Once $t$ is sufficiently large, by
far the most likely number of sampling events that will have taken
place will be equal to $t$.  Therefore one expects the properties of
the continuous and discrete time processes to concur at large times.
Additionally, we replace the transition probability from state $n'$ to
$n$ per generation $p_{n\, n'}$ with the transition rate per unit time
$T(n|n')$.

These two quantities are related via the expression
\begin{equation}
p_{n\,n'} =T(n|n') \Delta t + {\cal O} (\Delta t)^{2}\;,\quad n \ne n'
\label{p_to_T}
\end{equation}
where the terms of order $(\Delta t)^{2}$ and higher represent the
probability of two or more events happening in time $\Delta t$.
Substituting (\ref{p_to_T}) into Eq.~(\ref{markov_master}), and
letting $\Delta t \to 0$, gives
\begin{equation}
\frac{\rmd P(n, t)}{\rmd t} = \sum_{n' \neq n}\,T(n|n')\,P(n ',t) 
- \sum_{n' \neq n}\,T(n'|n)\,P(n,t)\,,
\label{master_usual} 
\end{equation}
which is the usual master equation used by statistical
physicists~\cite{vanKampen92,Gardiner04,Risken89}. Typically, the 
transition rates $T(n|n')$ are assumed given (this specifies the model) 
and we wish to determine the $P(n,t)$. The initial condition is that 
the system is in state $n_0$ at $t=0$.  Boundary conditions may also 
have to be given.

To illustrate these ideas we will look at three examples, all of which
were introduced in Section \ref{ideal}. We begin with the Moran model
for pure genetic drift, which has non-zero elements of the transition
matrix given by Eq.~(\ref{M_p_nonzero}).  In the continuous time
formulation, the population evolves by two individuals being sampled
uniformly with replacement from the population as a Poisson process
with the mean time between events taken to be unity.  One of the
sampled individuals is designated the parent which is copied to create
an offspring; the other sampled individual is sacrificed to make way
for the new offspring.  To obtain the transition rates in the master
equation (\ref{master_usual}), we need not consider the matrix element
$p_{n\,n}$, since this does not involve an actual transition.  We find
therefore
\begin{equation}
T(n+1|n) = \left( 1 - \frac{n}{N} \right) \left( \frac{n}{N}
\right)\,, \quad
T(n-1|n) = \left( \frac{n}{N} \right) \left( 1 -
\frac{n}{N} \right)\,.
\label{M_rates}
\end{equation}

We now turn to the Moran model with mutation, which has non-zero
elements given by Eq.~(\ref{M_p_nonzero_Mut}).  We will contrast two
ways one might construct a continuous-time process to mirror the
discrete time process as defined in Section~\ref{WF_M_mut}.  The most
obvious is to follow the prescription given earlier: that is, to have
Poisson sampling events as described in the previous paragraph, and to
mutate the copy of the parent to allele $B$ with probability $u$ if
the chosen parent was an $A$ or with probability $v$ in the opposite
direction.  The transition rates $T(n+1|n)$ and $T(n-1|n)$ are then
given by the expressions for $p_{n+1\,n}$ and $p_{n-1\,n}$ appearing
in Eq.~(\ref{M_p_nonzero_Mut}), i.e.,
\begin{eqnarray}
T(n+1|n) &=& (1-u) \left( 1 - \frac{n}{N} \right) \left( \frac{n}{N} \right)
+ v \left(1 - \frac{n}{N} \right)^{2}\,, \nonumber \\
T(n-1|n) &=& (1-v) \left( \frac{n}{N} \right) \left( 1 - \frac{n}{N} \right)
+ u \left( \frac{n}{N} \right)^{2}\,.
\label{M_Mut_rates_a}
\end{eqnarray}

The second approach we shall take differs in that the mutation process
will is divorced from the death/birth process---the two are considered
to take place independently with mutations occurring spontaneously,
rather than between a death and a birth.  Specifically, we shall
sample on average once per unit time a parent--offspring pair, as
previously described.  With probability $\gamma$, we replace the
offspring with a copy of the parent without mutation.  The rest of the
time, i.e., with probability $1-\gamma$, we switch the offspring
individual to allele state $B$ with probability $U$ if it is in
state $A$; otherwise we switch to state $A$ from $B$ with probability
$V$.  This leads to transition rates of the form
\begin{eqnarray}
T(n+1|n) &=& \gamma \left( 1 - \frac{n}{N} \right) \left( \frac{n}{N}
\right) + (1-\gamma) V \left( 1 - \frac{n}{N} \right) \\
T(n-1|n) &=& \gamma \left( \frac{n}{N}
\right) \left( 1 - \frac{n}{N} \right)  + (1-\gamma) U \left(
\frac{n}{N} \right) \;.
\label{M_Mut_rates_b}
\end{eqnarray}
To make contact with the discrete-time Moran process, we shall insist
that, at any given time, the probability that the \emph{next} event to
occur in the continuous-time process is equal to that for the
discrete-time process, given the same starting configuration.  To do
this, we rewrite $p_{n+1\,n}$ and $p_{n-1\,n}$ in
(\ref{M_p_nonzero_Mut}) as
\begin{eqnarray}
p_{n+1\, n} &=& (1-u-v) \left( 1 - \frac{n}{N} \right) \left(
\frac{n}{N} \right) + v \left( 1 - \frac{n}{N} \right) \nonumber\\
p_{n-1\, n} &=& (1-u-v) \left( 1 - \frac{n}{N} \right) \left(
\frac{n}{N} \right) + u \left( \frac{n}{N} \right) \;.
\label{M_p_nonzero_Mut_rates}
\end{eqnarray}
Correspondence is obtained if we put
\begin{equation}
\gamma = 1-u-v \;,\quad V = \frac{v}{u+v} \quad\mbox{and}\quad
U = \frac{u}{u+v} \;,
\label{connection_1}
\end{equation}
that is, if a fraction $u$ of the time, $A$ alleles spontaneously
change state to $B$ alleles, and a fraction $v$ of the time, $B$
alleles change to $A$.  Note this leads only to positive transition
rates if the mutation probabilities satisfy $u+v<1$.  Therefore, not
all Moran processes in which birth, death and mutation are intertwined
can be represented by a continuous-time model in which mutation occurs
independently of birth and death.  Nevertheless, a model with
transitions rates of this form was introduced by Maruyama (Model I in
the appendix of \cite{Maruyama77}) and it has been
remarked~\cite{Aalto89} that the results obtained by this model and
the Moran model seem to be essentially identical.  Our derivation of
the former starting from the latter shows why this is the case.

We now turn to the the Moran model with migration between two islands,
discussed in Section \ref{WF_M_mig}. In this case the state of the
system is specified by two integers: $\underline{n} = (n_{1},
n_{2})$. The considerations leading to the master equation
(\ref{master_usual}), where the state is specified by a single
variable, generalise in a straightforward way to the case where the
state is specified by a set of integers:
\begin{equation}
\frac{\rmd P(\underline{n}, t)}{\rmd t} = \sum_{\underline{n}' \neq
\underline{n}} \,T(\underline{n}|\underline{n} ')\,P(\underline{n}
',t) - \sum_{\underline{n}' \neq \underline{n}}\,T(\underline{n}
'|\underline{n}) \,P(\underline{n},t)\,,
\label{master_many} 
\end{equation}
where in the case under consideration here
$\underline{n}=(n_{1},n_{2})$. The transition probabilities are given
by Eq.~(\ref{M_mig_probs}).  To construct the corresponding
continuous-time process, we take the rate at which a parent is chosen
in deme $j$, the offspring of which displaces an individual in deme
$i$, to be
\begin{equation}
G_{ij} = f_j g_{ij} \;,
\label{migrates}
\end{equation}
where we recall  that $f_j$ is the probability the parent is in deme
$j$, and $g_{ij}$ the probability that the offspring lands in deme
$i$, conditioned on the parent being in deme $j$.  Since $\sum_j
f_j=1$ and $\sum_i g_{ij} = 1$ for any $j$, the total rate at which
pairs of individuals are sampled is
\begin{equation}
\sum_{i,j} G_{ij} = \sum_j f_j \sum_i g_{ij} = 1
\end{equation}
as required, even though the sampling is now non-uniform.

The transition probabilities for this model then follow as
\begin{eqnarray}
T(n_{1}+1,n_2|n_{1},n_2) &=& G_{11}  
\left( 1 - \frac{n_{1}}{N} \right) \frac{n_1}{N} + G_{1 2} 
\left( 1 - \frac{n_1}{N} \right) \frac{n_2}{N}\,, \nonumber \\
T(n_{1}-1,n_2|n_{1},n_2) &=& G_{11} 
\left( 1 - \frac{n_1}{N} \right) \frac{n_1}{N} + G_{1 2} 
\frac{n_1}{N} \left( 1 - \frac{n_2}{N} \right)\, \nonumber \\
T(n_{1},n_{2}+1|n_{1},n_2) &=& G_{22} 
\left( 1 - \frac{n_2}{N} \right) \frac{n_2}{N} + G_{2 1} 
\left( 1 - \frac{n_2}{N} \right) \frac{n_1}{N} \nonumber \\
T(n_{1},n_{2}-1|n_{1},n_2) &=& G_{22} 
\left( 1 - \frac{n_2}{N} \right) \frac{n_2}{N} + G_{2 1} 
\frac{n_2}{N} \left( 1 - \frac{n_1}{N} \right)\,.
\label{M_Mig_rates}
\end{eqnarray}
Note that the diagonal matrix elements $G_{ii}$ give the rates at
which an individuals are chosen to reproduce without their offspring
subsequently migrating.  From the master equation (\ref{master_many})
and the transition rates (\ref{M_Mig_rates}), we can obtain the
deterministic equation, valid in the limit $N \to \infty$, describing
the effect of migration on the number of $A$ alleles on one island, by
calculating $\rmd\langle n_{i} \rangle/\rmd t$. Focusing on island 1, we
multiply Eq.~(\ref{master_many}) by $n_{1}$ and sum over all $n_1$ and
$n_2$. By shifting the quantities being summed over by $\pm 1$ in some
of the sums over $n_1$, one finds that
\begin{equation}
\frac{\rmd\langle n_{1} \rangle}{\rmd t} = \langle T(n_{1}+1,n_2|n_{1},n_2) \rangle
- \langle T(n_{1}-1,n_2|n_{1},n_2) \rangle\,.
\label{av_mig1}
\end{equation}
Substituting for the transitions rates using Eq.~(\ref{M_Mig_rates})
yields
\begin{equation}
\frac{\rmd\langle n_{1} \rangle}{\rmd t} = G_{12} 
\left( \frac{\langle n_{2} \rangle}{N} - 
\frac{\langle n_{1} \rangle}{N} \right)\,,
\label{av_mig2}
\end{equation}
which is in agreement with Eq.~(\ref{M_Mig_deter}) as one would expect.

\subsection{Fokker-Planck equations and the large $N$ limit}
\label{FP_large_N}

Having obtained the transition rates (\ref{M_rates}), (\ref{M_Mut_rates_b})
and (\ref{M_Mig_rates}) for pure genetic drift involving two alleles, 
together with mutation and migration between two islands, we can now 
investigate the form that the governing equations take in the limit where $N$ 
is large. It will turn out that these processes can be described by 
Fokker-Planck equations in this limit. This will allow more complex 
situations, for example involving $L$ islands or $M$ alleles, to be formulated
in a rather straightforward way.
  
We again begin with the Moran model for pure genetic drift. When formulated as 
a master equation, it has the form (\ref{master_usual}) with transition 
rates (\ref{M_rates}). Such an equation is of ``the diffusion type'', since 
there is no macroscopic equation and the large $N$ approximation results in a 
non-linear Fokker-Planck equation~\cite{vanKampen92}. In this case, the large 
$N$ limit can be obtained in a straightforward fashion by substituting $n=xN$, 
and expanding out any remaining terms in $N$ in a $1/N$ expansion.
We find
\begin{eqnarray} 
\frac{\partial P}{\partial t} &=& \left( x - \frac{1}{N} \right) 
\left( 1-x+\frac{1}{N} \right) \left\{ P(x,t) - \frac{1}{N} 
\frac{\partial P(x, t)}{\partial x} + \frac{1}{2N^{2}} 
\frac{\partial^{2} P(x, t)}{\partial x^2} \right\} \nonumber \\
&+& \left( x + \frac{1}{N} \right) 
\left( 1-x-\frac{1}{N} \right) \left\{ P(x,t) + \frac{1}{N} 
\frac{\partial P(x, t)}{\partial x} + \frac{1}{2N^{2}} 
\frac{\partial^{2} P(x, t)}{\partial x^2} \right\} \nonumber \\
&-& 2 x(1-x) P(x, t) + {\cal O} (1/N^3) 
\label{M_2_largeN1}  
\end{eqnarray}
After a little algebra, this reduces to
\begin{equation} 
\frac{\partial P}{\partial t} = \frac{1}{N^2} 
\frac{\partial^{2}}{\partial x^2} \left[ x(1-x) P \right] 
+ {\cal O} (1/N^3)\,.
\label{M_2_largeN2}
\end{equation}
So if we define a scaled time $\tau = 2 t / N^{2}$, then taking $N \to
\infty$ we obtain
\begin{equation} 
\frac{\partial P}{\partial \tau} = \frac{1}{2} 
\frac{\partial^{2}}{\partial x^2} \left[ x(1-x) P \right]\,,
\label{diffusion_2_1}
\end{equation}
which is the diffusion equation (\ref{diffusion}).  This shows that
two different gene-based models---the Wright-Fisher and Moran
models---give the same mesoscopic Fokker-Planck equation. This is a
familiar situation in statistical physics: microscopic models
describing the same ``physics'', but formulated in slightly different
ways, typically will give rise to the same mesoscopic equation.  The
only difference between the Wright-Fisher and the Moran model is that,
as usual, the time-scales differ. The redefinition of a generation in
the Wright-Fisher model leads to $\tau'=t/N$, whereas the scaling of
time in the Moran model gives $\tau= 2t/N^{2}$. In other words, a
generation in the Wright-Fisher model is $N/2$ times as long as in the
Moran model in this mesoscopic description.

When the diffusion approximation is applied to the Moran model which
includes mutational and migrational effects, the rates of migration
and mutation have to be rescaled by a factor of $N$. If this were not
so, the large $N$ limit of the master equation would not be a
Fokker-Planck equation of the diffusion type, but a linear
Fokker-Planck equation which would describe fluctuations about the
deterministic equation (\ref{WF_Mut_deter})~\cite{vanKampen92}.  In
the backward-time formulation discussed later (see
Section~\ref{coalescent}), we will also see that this scaling
corresponds to a regime in which all relevant processes affecting the
ancestry of the population occur on the same timescale.

Let us first consider the models with mutation.  The scaled mutation
rates, with the factor of $N$ mentioned above and a factor of 2 to be
explained below, are defined by
\begin{equation}
{\cal U} =  \frac{u N}{2}\,, \ \ \ \ {\cal V} = \frac{v N}{2}\,.
\label{scaled_M_rates}
\end{equation}
In Section \ref{continuous_time} we discussed two ways of introducing
mutation into the Moran model. The first resulting in transition rates
given by Eq.~(\ref{M_Mut_rates_a}) and the second in transition rates
given by Eq.~(\ref{M_Mut_rates_b}), with $\gamma, U$ and $V$ given by
Eq.~(\ref{connection_1}). There is no need to carry out the $1/N$
expansion for each model in turn, since the transitions rates for both
are identical (although one has the restriction $u+v<1$), and which
using the scaled forms (\ref{scaled_M_rates}) read
\begin{eqnarray}
T(n+1|n) &=& \left(1-\frac{2{\cal U}}{N}-\frac{2{\cal V}}{N} \right) 
\left( 1 - \frac{n}{N} \right) \left( \frac{n}{N} \right) 
+ \frac{2{\cal V}}{N} \left( 1 - \frac{n}{N} \right) \nonumber \\
T(n-1|n) &=& \left(1-\frac{2{\cal U}}{N}-\frac{2{\cal V}}{N} \right) 
\left( \frac{n}{N} \right) \left( 1 - \frac{n}{N} \right)  
+ \frac{2{\cal U}}{N} \left(\frac{n}{N} \right) \;.
\label{M_Mut_rates_c}
\end{eqnarray}
Since the term (\ref{M_2_largeN2}) is already of order $1/N^{2}$, the terms 
involving ${\cal U}/N$ and ${\cal V}/N$ which multiply it give terms of order
$1/N^{3}$, which do not contribute to the Fokker-Planck equation. Therefore 
the only additional terms due to mutation are the second terms on the 
right-hand side of the transitions probabilities (\ref{M_Mut_rates_c}), which 
are
\begin{eqnarray} 
&&\left\{ \frac{2{\cal V}}{N} \left( 1-x+\frac{1}{N} \right) \right\} 
\left\{ P(x,t) - \frac{1}{N} \frac{\partial P(x, t)}{\partial x}  \right\} 
+ \left\{ \frac{2{\cal U}}{N} \left( x + \frac{1}{N} \right) \right\} 
\nonumber \\
&\times& \left\{ P(x,t) + \frac{1}{N} \frac{\partial P(x, t)}{\partial x}  
\right\} - \left[ \frac{2{\cal V}}{N} (1-x) + \frac{2{\cal U}}{N} x \right] 
P(x, t) + {\cal O} (1/N^3)\,,
\label{M_2_Mut1}  
\end{eqnarray}
which on simplification yields
\begin{equation}
\frac{2}{N^2} \frac{\partial}{\partial x} 
\left[ \left\{ {\cal U}x - {\cal V}(1-x) \right\} P \right] + 
{\cal O} (1/N^3)\,.
\label{M_2_Mut2}
\end{equation}
Adding this additional term to Eq.~(\ref{M_2_largeN2}), and using rescaled 
time units $\tau=2t/N^2$, we find on letting $N \to \infty$ that the 
additional mutation processes present in either model give rise to the 
Fokker-Planck equation
\begin{equation}
\frac{\partial P}{\partial \tau} =
\frac{\partial}{\partial x} \left[ \left\{ {\cal U}x - {\cal V}(1-x)
\right\} P \right] + \frac{1}{2} \frac{\partial^{2}}{\partial x^2}
\left[ x(1-x) P \right]\,.
\label{diffusion_2_1_Mut}
\end{equation}
The restriction $u+v<1$, which applied to one of the models, takes the form 
${\cal U} + {\cal V} <N/2$, when expressed in terms of the scaled mutation 
rates. Since in most applications ${\cal U}$ and ${\cal V}$ are numerically 
small, and in any case we are taking the limit $N \to \infty$, this condition 
will always hold. Furthermore, Fokker-Planck equation (\ref{diffusion_2_1_Mut})
agrees with that found by applying the diffusion approximation to the 
Wright-Fisher model with mutation~\cite{Crow70}. The additional factors of 2 
introduced in Eq.~(\ref{scaled_M_rates}) were included to conform with the 
conventions used in the Wright-Fisher case.

The introduction of two islands or speakers has been discussed in
section \ref{WF_M_mig} and the transition rates given in
Eq.~(\ref{M_Mig_rates}). To derive the Fokker-Planck equation from the
master equation in this case, we note that the first terms on the
right-hand sides of these transition rates are exactly as for pure
drift, up to the inclusion of prefactors $G_{11}$ or $G_{22}$.
Therefore, the contribution from events which do not involve actual
migration is
\begin{equation} 
\sum^{2}_{i=1} \frac{G_{ii}}{N^2} \frac{\partial^{2}}{\partial x_{i}^2} 
\left[ x_{i}(1-x_{i}) P \right] + {\cal O} (1/N^3)\,.
\label{M_2_largeN_2_1}
\end{equation}
The second terms on the right-hand side of the transition rates in 
Eq.~(\ref{M_Mig_rates}), for migration from island $j$ to island $i$ are:
\begin{eqnarray} 
&+& G_{i j} \left( 1 - x_{i} + \frac{1}{N} \right) x_{j} 
\left\{ P(\underline{x},t) - \frac{1}{N} \frac{\partial P(\underline{x}, t)}
{\partial x_i} \right\} \nonumber \\
&+& G_{i j} \left( x_{i} + \frac{1}{N} \right) \left( 1-x_{j} \right) 
\left\{ P(\underline{x},t) + \frac{1}{N} \frac{\partial P(\underline{x}, t)}
{\partial x_i} \right\} \nonumber \\
&-& G_{i j} \left[ \left( 1 - x_{i} \right) x_{j} + x_{i} 
\left( 1 - x_{j} \right) \right] P(\underline{x},t) 
+ {\cal O} \left( \frac{1}{N^2}\right) \nonumber \\
&=& G_{i j} \frac{1}{N}\,\frac{\partial }
{\partial x_{i}} \left[ \left( x_{i} - x_{j} \right) P \right] 
+ {\cal O} \left( \frac{1}{N^2}\right)\,.
\label{M_2_largeN_2_2}
\end{eqnarray} 
In order for the Fokker-Planck equation to be of the diffusion type,
the rates of migration between demes, i.e., $G_{ij}$ for $i\ne j$,
have to be scaled by a factor of $N$, just as the mutation rates were.
Let us now parametrise these off-diagonal matrix elements as $G_{ij} =
2 {\cal G}_{ij}/N$ where ${\cal G}_{ij}$ is of order unity.  Then,
$G_{ii} = f_i + {\cal O}(1/N)$, and so the Fokker-Planck equation becomes
\begin{eqnarray}
\frac{\partial P(\underline{x},t)}{\partial t} &=&
\frac{1}{N^2} \sum_{i=1}^{2} f_i \frac{\partial^{2}}{\partial x_{i}^2} 
\left[ x_{i}(1-x_{i}) P \right] \nonumber \\
&&{}+ \frac{2}{N^2} \left( {\cal G}_{1 2} \frac{\partial }{\partial x_{1}} - 
{\cal G}_{2 1} \frac{\partial } {\partial x_{2}} \right) 
\left[ \left( x_{1} - x_{2} \right) P \right] + 
{\cal O} \left( \frac{1}{N^3}\right)\,.
\label{M_2_largeN_2_3}
\end{eqnarray}
Introducing $\tau = 2t / N^{2}$, as before, we can let $N \to \infty$ and 
obtain the Fokker-Planck equation for migration between two islands:
\begin{eqnarray}
\frac{\partial P}{\partial \tau} &=& \frac{1}{2} \sum_{i=1}^{2} f_i
\frac{\partial^{2}} {\partial x_{i}^2} \left[ x_{i}(1-x_{i}) P \right]
\nonumber \\ 
&+& \left( {\cal G}_{1 2} \frac{\partial }{\partial x_{1}} - {\cal G}_{2 1} 
\frac{\partial }{\partial x_{2}} \right) \left[ \left( x_{1} - x_{2} \right) 
P \right]\,. 
\label{diffusion_2_2_Mut}
\end{eqnarray} 
In this mesoscopic formulation all trace of the individual genes has
disappeared, and we are left only with a description in terms of the
fractions of genes on island $i$ which are of type $A$.

\subsection{Migration of genes having $M$ alleles between $L$ islands}
\label{M_and_L}

We have so far, for simplicity, mainly restricted ourselves to considering two
islands and two alleles. However the formalism we have discussed carries over
with little modification to the cases of an arbitrary number of island and 
alleles. In order to avoid introducing too many complicating looking 
expressions in this section, we have relegated some of the details to an 
appendix.

\subsubsection{L Islands}
\label{L_islands}

We assume that on each island we have a deme labelled by $i$, with
$n_{i}$ being the number of $A$ alleles and $(N-n_{i})$ the number of
$B$ alleles. A parent is chosen from island $j$ with probability $f_j$
and subsequently leaves an offspring on island $i$ with probability
$g_{ij}$, where the sum rule $g_{ii} = 1 - \sum_{j\ne i} g_{ji}$ must
be satisfied. Then proceeding as in Section \ref{WF_M_mig}, the
transition probabilities analogous to those in Eq.~(\ref{M_mig_probs})
are:
\begin{eqnarray}
p_{(n_{1},\ldots,n_{i}+1,\ldots,n_L)\,(n_{i},\ldots,n_L)} &=& 
f_i \left( 1 - \sum_{j \neq i} g_{j i} \right) 
\left( 1 - \frac{n_{i}}{N} \right) \frac{n_i}{N} \nonumber \\
&+& \sum_{j \neq i} f_j g_{i j} 
\left( 1 - \frac{n_i}{N} \right) \frac{n_j}{N}\,, 
\label{M_mig_probs_L+}
\end{eqnarray}
and
\begin{eqnarray}
p_{(n_{1},\ldots,n_{i}-1,\ldots,n_L)\,(n_{i},\ldots,n_L)} &=& 
f_i \left( 1 - \sum_{j \neq i} g_{j i} \right) 
\left( 1 - \frac{n_{i}}{N} \right) \frac{n_i}{N} \nonumber \\
&+& \sum_{j \neq i} f_j g_{i j} 
\frac{n_i}{N} \left( 1 - \frac{n_j}{N} \right)\,.
\label{M_mig_probs_L-}
\end{eqnarray}
To go over to a master equation description, we introduce the set of
migration rates via Eq.~(\ref{migrates}).  The transition rates for
the master equation (\ref{master_many}) now follow:
\begin{eqnarray}
T(n_{1},\ldots,n_{i}+1,\ldots,n_L|n_{1},\ldots,n_L) &=& 
\sum_{j} G_{ij} \left( 1 - \frac{n_i}{N} \right) 
\frac{n_j}{N}\,, \nonumber \\
T(n_{1},\ldots,n_{i}-1,\ldots,n_L|n_{1},\ldots,n_L) &=& 
\sum_{j} G_{i j} \frac{n_i}{N} \left( 1 - \frac{n_j}{N} \right)\,.
\label{M_mig_rates_L}
\end{eqnarray}
To obtain the Fokker-Planck equation, valid when $N$ is large, is now
very simple, since the required calculations are as in the $L=2$
case. For example, Eq.~(\ref{M_2_largeN_2_1}) is unchanged, except
that the sum is now from $i=1$ to $i=L$, and the contribution
(\ref{M_2_largeN_2_2}) holds for all pairs $i$ and $j$ with $i \neq
j$. Introducing as before $G_{i j} = 2 {\cal G}_{i j}/N$ for $i\ne j$,
rescaling time as $\tau = 2t/N^{2}$, letting $N \to \infty$ and noting
that $G_{ii}\to f_i$ in this limit, we obtain the Fokker-Planck equation
for migration between $L$ islands:
\begin{eqnarray}
\frac{\partial P}{\partial \tau} &=& \frac{1}{2} \sum_{i=1}^{L} f_i
\frac{\partial^{2}} {\partial x_{i}^2} \left[ x_{i}(1-x_{i}) P \right] 
\nonumber \\
&+& \sum_{\langle i j \rangle} \left( {\cal G}_{i j} \frac{\partial }
{\partial x_{i}} - {\cal G}_{j i} \frac{\partial }{\partial x_{j}} \right) 
\left[ \left( x_{i} - x_{j} \right) P \right]\,. 
\label{diffusion_2_L_Mut}
\end{eqnarray} 
Here the notation $\langle i j \rangle$ means ``sum over all distinct pairs
$i$ and $j$''. 

Fokker-Planck equations may be written as continuity equations~\cite{Risken89};
in the case of interest here it takes the form $\partial P/\partial \tau +
\sum_{i} \partial J_{i}/\partial x_{i} = 0$, where
\begin{eqnarray}
J_{i} (\underline{x}, t) &=& - \frac{1}{2} f_i \frac{\partial}{\partial x_{i}} 
\left[ x_{i} (1-x_{i}) P (\underline{x}, t) \right] 
- \sum_{j \neq i} {\cal G}_{i j} \left[ \left( x_{i} - x_{j} \right) 
P (\underline{x}, t) \right]\,, 
\label{current}
\end{eqnarray}
is the probability current. Multiplying the equation of continuity by $x_k$ 
and integrating over all $x_i$ ($i=1,\ldots,L$) gives
\begin{equation}
\frac{\rmd\langle x_{k} \rangle}{\rmd\tau} = \int d\underline{x} x_{k} 
\frac{\partial P}{\partial \tau} =  - \sum_{i} \int d\underline{x} x_{k} 
\frac{\partial J_i}{\partial x_i} = \int d\underline{x} J_{k}\,, 
\label{av_xk_1}
\end{equation}
since the probability current vanishes on the boundaries~\cite{Risken89}.
From Eqs.~(\ref{current}) and (\ref{av_xk_1}) we have
\begin{equation}
\frac{\rmd \langle x_{k} \rangle}{\rmd\tau} = - \sum_{j \neq k} {\cal G}_{kj} 
\left( \langle x_{k} \rangle - \langle x_{j} \rangle \right)\,. 
\label{averages_2_L}
\end{equation}
To make contact with previous results for finite $N$, we reintroduce 
$n_{i}=Nx_i$, $G_{i j} = 2 {\cal G}_{i j}/N$ and $t=N^{2}\tau/2$, so that 
Eq.~(\ref{averages_2_L}) becomes
\begin{equation}
\frac{\rmd \langle n_{k} \rangle}{\rmd t} = 
- \sum_{j \neq k} G_{k j} \frac{1}{N} \left( \langle n_{k} \rangle 
- \langle n_{j} \rangle \right)\,. 
\label{averages_2_L_scaled}
\end{equation}
This is in agreement with Eq.~(\ref{av_mig2}) found directly from the
master equation in the case $L=2$.

Let us briefly consider under what conditions the mean of the frequency of 
$A$ alleles across all islands is conserved.  For this we require
\begin{equation}
\frac{\rmd}{\rmd t} \sum_k \langle n_k \rangle = - \sum_{k} \sum_{j
\ne k} \frac{G_{kj}}{N} \left( \langle n_{k} \rangle 
- \langle n_{j} \rangle \right) = \sum_{j} \frac{\langle n_j \rangle}{N} 
\sum_{k \ne j} \left( G_{kj} - G_{jk} \right) = 0 \;.
\end{equation}
For this to hold for any set of allele frequencies $\langle n_j
\rangle$, we require
\begin{equation}
\label{consG}
\sum_{k \ne j} G_{kj} = \sum_{k \ne j} G_{jk} \quad \forall\, j \;.
\end{equation}
Since the total mean $A$ allele frequency is conserved by migration
(and rigorously conserved in the absence of drift, not just on
average) models for which the migration rates satisfy this equality
are called \emph{conservative}.  We will see in Section~\ref{nonideal}
that a number of exact results are known for models with conservative
migration.

\subsubsection{$M$ Alleles}
\label{M_alleles}

So far in this article we have restricted ourselves to situations
where each gene has only two alleles: $A$ and $B$. There is no reason,
other than grounds of simplicity, to do so, and it is possible to
generalise the treatment to the situation where there are $M$ possible
types of alleles, $A_{\alpha}$, where $\alpha=1,\ldots,M$. If one is
considering a single island, the possible states are labelled by the
number of $A_{\alpha}$ alleles denoted by $n_{\alpha}$, where
$\alpha=1,\ldots,(M-1)$. Clearly, since $\sum^{M}_{\alpha=1}
n_{\alpha} = 1$, $n_{M}= 1 - \sum^{(M-1)}_{\alpha=1} n_{\alpha}$ is
not an independent variable.  If there are $L$ islands, then there are
$L(M-1)$ independent variables: $n_{i \alpha}$ with $i=1,\ldots,L$ and
$\alpha=1,\ldots,(M-1)$.

We begin by assuming that there is no mutation or migration (that is, $L=1$).
The basic transition probability associated with an $A_{\beta}$ allele being
replaced by an $A_{\alpha}$ allele is 
\begin{equation}
p_{(n_{1},\ldots,n_{\alpha}+1,\ldots,n_{\beta}-1,\ldots,n_{M-1})\,
(\underline{n})} = \left( \frac{n_\alpha}{N} \right) 
\left( \frac{n_\beta}{N} \right)\,,
\label{trans_probs_M_1_1}
\end{equation}
where $\alpha \neq \beta$, $\alpha,\beta=1,\ldots,(M-1)$. Clearly if
the transition involves the $M$th allele the expression has to be
modified, since the number of alleles of type $M$ is not an
independent variable.  The transition rates which appear in
the master equation are obtained as before through consideration of a
continuous-time process in which pairs of individuals, one of which
reproduces and whose offspring displaces the others, are sampled on
average once per unit time.  This leads to
\begin{equation}
T(n_{1},\ldots,n_{\alpha}+1,\ldots,n_{\beta}-1,\ldots,n_{M-1}|\underline{n})
= \left( \frac{n_{\alpha}}{N} \right) \left( \frac{n_{\beta}}{N} \right)\,, 
\label{rates_M_1_1}
\end{equation}
where $\alpha \neq \beta$ and neither is equal to $M$, and
\begin{eqnarray}
T(n_{1},\ldots,n_{\alpha}+1,\ldots,n_{M-1}|\underline{n}) &=& 
T(n_{1},\ldots,n_{\alpha}-1,\ldots,n_{M-1}|\underline{n}) \nonumber \\
&=& \left( \frac{n_{\alpha}}{N} \right)\,
\left( \frac{N - \sum_{\beta=1}^{M-1} n_{\beta}}{N} \right) \,,
\label{rates_M_1_2}
\end{eqnarray}
if either allele $\alpha\,(\neq M)$ increases at the expense of allele $M$
or allele $M$ increases at the expense of allele $\alpha\,(\neq M)$, 
respectively. 

We can now carry out a large $N$ expansion on the master equation with
the transition rates (\ref{rates_M_1_1}) and (\ref{rates_M_1_2}), just
as we did in Eq.~(\ref{M_2_largeN1}). The algebraic details are a
little more messy for general $M$, and so we have given them in the
Appendix. Rescaling the time as before, one finds the Fokker-Planck
equation
\begin{equation}
\frac{\partial P}{\partial \tau} = \frac{1}{2} \left\{ \sum^{M-1}_{\alpha=1} 
\frac{\partial^{2} }{\partial x_{\alpha}^{2}} \left[ x_{\alpha} 
\left( 1 - x_{\alpha} \right) P \right] - \sum^{M-1}_{\alpha=1} 
\sum^{M-1}_{\beta \neq \alpha} \frac{\partial^{2} }{\partial x_{\alpha} 
\partial x_{\beta}} \left[ x_{\alpha} x_{\beta} P \right] \right\}\,,
\label{diffusion_M_1}
\end{equation}
in the limit $N \to \infty$. This the usual multi-allelic diffusion
equation~\cite{Crow56}

If we now add the possibility of mutations occurring, $M(M-1)$
mutation probabilities have to be defined to describe mutations from
allele $A_\beta$ to allele $A_\alpha$. We will denote this probability
by $u_{\alpha \beta}\,(\alpha \neq \beta)$. To see how this matrix
enters the formalism, let us return to the the case of two alleles and
rewrite Eqs.~(\ref{Mut_p1}) and (\ref{Mut_p2}) as 
\begin{equation}
p_{\alpha} = \sum^{2}_{\beta=1} u_{\alpha \beta} (n_{\beta}/N)\,,
\label{p_alpha}
\end{equation}
where we have written $A$ and $B$ as $A_1$ and $A_2$ respectively, 
$n_{2}=N-n_{1}$, and where the \textit{mutation matrix} is given by
\begin{eqnarray}
\label{Mut_matrix}
{\bf u}=\left[\matrix{1-u&v\cr
                u&1-v\cr}\right]\,. 
\end{eqnarray}
Notice that the condition that the columns of the matrix ${\bf u}$ add up to
1 ensures that $\sum_{\alpha} p_{\alpha} = 1$, since
$\sum_{\beta} n_{\beta} = N$. For the case of $M$ different alleles ${\bf u}$ 
has entries $u_{\alpha \beta}$ if $\alpha \neq \beta$, and 
$1-\sum_{\gamma \neq \beta} u_{\gamma \beta}$ for the diagonal entry 
$(\beta,\beta)$. Therefore to deduce the form of the transition probabilities 
in the Moran model with $M$ alleles and mutation, we may follow the same 
arguments which led to Eq.~(\ref{M_p_nonzero_Mut}) in the case of two alleles. 
For instance, if an $A_\beta$ allele is replaced by an $A_\alpha$ allele, then 
\begin{equation}
p_{(n_{1},\ldots,n_{\alpha}+1,\ldots,n_{\beta}-1,\ldots,n_{M-1})\,
(\underline{n})} = p_{\alpha} (\underline{n}) 
\left( \frac{n_{\beta}}{N} \right)\,.
\label{trans_probs_M_Mut_1}
\end{equation}
This will hold for all $\alpha$ and $\beta$, with $\alpha \neq \beta$, as
long as we interpret $n_{M}$ as $1 - \sum^{(M-1)}_{\gamma=1} n_{\gamma}$. The
change in the mean value of $n_{\alpha}$ with time can be calculated in an 
analogous way to Eq.~(\ref{M_Mut_deter}) and one finds that
\begin{equation}
\langle n_{\alpha}(t+1) \rangle = \langle n_{\alpha}(t) \rangle + 
\sum_{\beta \neq \alpha}  \left\{ u_{\alpha \beta} 
\frac{\langle n_{\beta}(t) \rangle}{N} - u_{\beta \alpha} 
\frac{\langle n_{\alpha}(t) \rangle}{N} \right\}\,.
\label{M_Mut_M_deter}
\end{equation}

As in Section~\ref{continuous_time}, we shall follow two approaches to
mutation in the continuous time formulation.  First, we shall have
pairs of individuals sampled on average once per unit time, and the
offspring of the parent changed to allele $\alpha$ with probability
$u_{\alpha \beta}$ given that the sampled parent carried allele
$\beta$.  Transition rates for this model are then given by
\begin{equation}
T(n_1 \ldots n_{\alpha}+1 \ldots n_{\beta}-1 \ldots n_{M-1}|
\underline{n}) = \left(
\sum_{\delta} u_{\alpha \delta} \frac{n_{\delta}}{N} \right)
\frac{n_{\beta}}{N}
\label{muta}
\end{equation}
for $\alpha\neq\beta$ and it is understood that $n_{M}$ is always
implicitly given by $1 - \sum_{\alpha=1}^{M-1} n_{\alpha}$. In
the Appendix we derive the Fokker-Plank equation for this system,
valid when $N$ is large.  Again, it is necessary to rescale the rates
of mutation between different alleles, i.e.,
\begin{equation}
{\cal U}_{\alpha \beta} = \frac{N u_{\alpha \beta}}{2} \quad \alpha
\neq \beta \;,
\label{scaled_rates_general}
\end{equation}
The Fokker-Planck equation reads
\begin{equation}
\frac{\partial P}{\partial \tau} = - \sum^{M-1}_{\alpha=1} 
\frac{\partial }{\partial x_{\alpha}} \left[ {\cal A}_{\alpha} (\underline{x}) 
P \right] + \frac{1}{2} \sum^{M-1}_{\alpha=1} \sum^{M-1}_{\beta=1}
\frac{\partial^{2} }{\partial x_{\alpha} \partial x_{\beta}} 
\left[ {\cal D}_{\alpha \beta}(\underline{x}) P \right] \,,
\label{diffusion_M_Mut_1}
\end{equation}
where ${\cal A}_{\alpha} (\underline{x})$ and 
${\cal D}_{\alpha \beta} (\underline{x})$ are given by 
\begin{equation}
{\cal A}_{\alpha} (\underline{x}) = \sum_{\beta=1}^{M} \left({\cal
U}_{\alpha \beta} x_{\beta} - {\cal U}_{\beta \alpha} x_{\alpha} \right) \,,
\label{A_M_1}
\end{equation}
and
\begin{equation}
{\cal D}_{\alpha \beta} (\underline{x}) = \left\{ \begin{array}{ll} 
x_{\alpha} (1-x_{\alpha}) , & \mbox{\ if $\alpha=\beta$} \\
- x_{\alpha}x_{\beta} , & \mbox{\ if $\alpha \neq \beta$\,.} 
\end{array} \right.
\label{D_M_1}
\end{equation}
Note that the term in the sum with $\alpha=\beta$ in (\ref{A_M_1}) is
zero, so it is not necessary to define the diagonal rescaled mutation
matrix elements ${\cal U}_{\alpha \alpha}$

It is worth, briefly, considering the variant of the continuous-time
mutation model in which mutation events are separated from the
birth-death events.  As in Section~\ref{continuous_time}, we shall
sample pairs of individuals on average once per generation, and a
fraction $\gamma$ of the time replace one of the pair with a copy of
the other; the remaining fraction $1-\gamma$ of the time, one of the
pair shall be mutated to allele $\alpha$ with probability $U_{\alpha
  \beta}$, given that it carries allele $\beta$.  The transition rates
for this model are
\begin{equation}
\label{mutb}
\fl T(n_1 \ldots n_{\alpha}+1 \ldots n_{\beta}-1 \ldots n_{M-1}|
(\underline{n}) = \gamma
\left[ \frac{n_{\alpha}}{N} \frac{n_{\beta}}{N} \right] + (1-\gamma)
\left[ U_{\alpha \beta} \frac{n_{\beta}}{N} \right] \;.
\end{equation}
Compare now with (\ref{muta}) which can be written as
\begin{equation}
\fl T(n_1 \ldots n_{\alpha}+1 \ldots n_{\beta}-1 \ldots n_{M-1}|
\underline{n}) = \left( 1 - \sum_{\delta \ne \alpha} u_{\delta
\alpha} \right) \frac{n_{\alpha}}{N}
\frac{n_{\beta}}{N} + \sum_{\delta \ne \alpha} u_{\alpha \delta}
\frac{n_{\delta}}{N} \frac{n_{\beta}}{N} \;.
\end{equation}
Since no allele frequencies $n_{\delta}/N$ with $\delta \ne \alpha,
\beta$ appear in (\ref{mutb}), the only way the two can be made equal
is if we take $u_{\alpha \delta} = u_{\alpha}$. Then, the sum over
$n_{\delta}/N$ gives $1-n_{\alpha}/N$, since all allele frequencies
add up to unity.  This results in
\begin{equation}
T(n_1 \ldots n_{\alpha}+1 \ldots n_{\beta}-1 \ldots n_{M-1}|
\underline{n}) = \left( 1 - \sum_{\delta} u_{\delta} \right)
\frac{n_{\alpha}}{N} \frac{n_{\beta}}{N} + u_{\alpha}
\frac{n_{\beta}}{N} \;.
\label{mut_second}
\end{equation}
Correspondence with (\ref{mutb}) is then obtained as long as
\begin{equation}
\gamma = 1 - \sum_{\delta} u_{\delta} \quad\mbox{and}\quad
U_{\alpha} = \frac{u_{\alpha}}{\sum_{\delta} u_{\delta}}\,,
\end{equation}
where we have taken $U_{\alpha \beta} = U_{\alpha}$ for all $\beta \ne
\alpha$.  We see that, in common with the case $M=2$ discussed in
Section~\ref{continuous_time}, the sum of mutation rates
$\sum_{\delta} u_{\delta}$ must be less than one in order for the
simplified model (in which mutation is a spontaneous process) to have
the same dynamics as the full model in which mutation, birth and death
are combined.

We remark that this condition is satisfied in the limit $N\to\infty$
when mutation rates are scaled with $N$ as above.  The Fokker-Planck
equation (\ref{diffusion_M_Mut_1}) results, albeit with a simplified
deterministic (potential) term
\begin{equation}
{\cal A}_{\alpha} (\underline{x}) = {\cal U}_{\alpha} - \sum_{\beta=1}^{M} 
{\cal U}_{\beta} x_{\alpha} \,,
\label{cal_A_mod}
\end{equation}
on account of the mutation rates depending only on the end product of
the mutation. Note that this time, the term $\beta=\alpha$ is
included in the summation.  Even with this simplification, the
Fokker-Planck equation is a nonlinear partial differential in $(M-1)$
variables, and one therefore does not expect it to be readily solved.
The stationary solution has long been known \cite{Wright45} to be
\begin{equation}
\label{multiPast}
P^\ast(x_1, x_2, \ldots, x_M) = \Gamma \left(2 \sum_{\alpha=1}^{M} {\cal
U}_{\alpha} \right) \prod_{\alpha=1}^{M} \frac{x_{\alpha}^{2 {\cal U}_\alpha -
1}}{\Gamma(2 {\cal U}_\alpha)}
\end{equation}
in which $\Gamma(u)$ is the usual Gamma function.  In this steady
state, the probability current vanishes everywhere, i.e., detailed
balance is satisfied.  For a general set of mutation rates, i.e., one
for which the mutation rates depend on both initial and final allele
states, the stationary solution of the Fokker-Planck equation has not
been found.  One therefore speculates that detailed balance is not
satisfied for these more general models, which may be related to the
fact that one cannot represent these mutation rates in models where
birth/death and mutation are independent.

Meanwhile, a remarkable feature of the Fokker-Planck equation under
the restriction ${\cal U}_{\alpha \beta} = {\cal U}_{\alpha}$ is that
the full time dependence can be found, not just the steady state.
There are at least two ways to achieve this; either one can explicitly
construct the set of polynomials orthogonal to the weight function
given by (\ref{multiPast}) to obtain the eigenfunctions of the
operator appearing in (\ref{diffusion_M_Mut_1}) \cite{Griffiths79}.
Alternatively, one can make the change of variables 
\begin{equation}
\xi_{\alpha} = \frac{x_{\alpha}}{1 - \sum_{\beta < \alpha} x_{\beta}}\,, \ 
\alpha=1,\ldots,M-1\,,
\label{xis}
\end{equation}
under which the equation becomes separable~\cite{Baxter07}.
Essentially this is possible because of the simple way that the
$M$-allele problem is ``nested'' in the $(M+1)$-allele problem.  This
is also responsible for most of the simplifications in the
calculations of many other quantities of interest, such as mean times
to fixation, the probability of a particular sequence of extinctions,
and so on~\cite{Baxter07}.  In this approach, it is also learnt that
the eigenfunctions of the Fokker-Planck equation are Jacobi polynomials.

\subsubsection{$L$ Islands and $M$ Alleles}
\label{L_and_M}

The final case to consider is when there are $M$ alleles and $L$ islands. 
The analysis is a combination of the $M=2$, general $L$ case considered in 
section \ref{M_alleles} and the $L=1$, general $M$ case just considered. So,
for example, not including mutation for the moment, the terms in the transition
probability which do not involve migration are exactly as in 
Eq.~(\ref{rates_M_1_1}) with indices $i$ added to the $n_{\alpha}$ or 
$n_{\beta}$, as well as an extra factor of $f_i (1-\sum_{j \neq i} g_{ij})$.
Alternatively, the transition probabilities are exactly as in 
Eqs.~(\ref{M_mig_probs_L+}) and (\ref{M_mig_probs_L-}) with indices $\alpha$
and $\beta$ added:
\begin{eqnarray}
p_{(\ldots,n_{i \alpha}+1,\ldots,n_{i \beta}-1,\ldots)\,(\underline{n})} 
&=& f_i \left( 1 - \sum_{j \neq i} g_{j i} \right) 
\left( \frac{n_{i \beta}}{N} \right) \left( \frac{n_{i \alpha}}{N} \right)
\nonumber \\
&+& \sum_{j \neq i} f_j g_{i j} 
\left( \frac{n_{i \beta}}{N} \right) \left( \frac{n_{j \alpha}}{N} \right)\,. 
\label{M_M_mig_probs_L}
\end{eqnarray}
Here we have removed an $A_{\beta}$ allele from island $i$ and replaced it 
with an $A_{\alpha}$ allele, also from island $i$ (first term) or with a 
migrant from island $j$ (second term).  

We assume that mutation occurs at the point that a parent is copied to
make the new offspring, no matter whether that offspring remains on
its parent's island or emigrates.  Furthermore, the probability that
a parent with allele $\beta$ gives birth to an offspring with allele
$\alpha$ does not depend on either the source or target islands, i.e.,
the mutation occurs with probability $u_{\alpha \beta}$ as before.
Including this mutation process, the transition probability becomes
\begin{eqnarray}
p_{(\ldots,n_{i \alpha}+1,\ldots,n_{i \beta}-1,\ldots)\,(\underline{n})} 
&=& f_i \left( 1 - \sum_{j \neq i} g_{j i} \right) 
p_{\alpha} (\underline{n}_{i}) \left( \frac{n_{i \beta}}{N} \right)
\nonumber \\
&+& \sum_{j \neq i} f_j g_{i j} 
p_{\alpha} (\underline{n}_{j}) \left( \frac{n_{i \beta}}{N} \right)\,. 
\label{M_M_Mut_mig_probs_L}
\end{eqnarray}
As before, $\alpha \neq \beta$ and it is understood that $n_{i M}$ should be 
replaced by $1 - \sum^{(M-1)}_{\gamma=1} n_{i \gamma}$.  The
transition rates follow as before.

Starting from the master equation (\ref{master_many}) it is
straightforward to obtain the Fokker-Planck equation using these
transition rates, the algebraic steps simply being a combination of
those which need to be carried out for the cases considered in
Sections \ref{L_islands} and \ref{M_alleles}.  The details are given
in the Appendix, where it is shown that the resulting
Fokker-Planck equation is given by
\begin{eqnarray}
\frac{\partial P}{\partial \tau} &=& - \sum^{L}_{i=1} f_i
\sum^{M-1}_{\alpha=1} \frac{\partial }{\partial x_{i \alpha}} 
\left[ {\cal A}_{\alpha} (\underline{x}_i) P \right] \nonumber \\
&+& \sum_{\langle i j \rangle} \sum^{M-1}_{\alpha=1} 
\left( {\cal G}_{i j} \frac{\partial }{\partial x_{i \alpha}} - 
{\cal G}_{j i}\frac{\partial }{\partial x_{j \alpha}} \right) 
\left[ \left( x_{i \alpha} - x_{j \alpha} \right) P \right]
\nonumber \\
&+& \frac{1}{2} \sum^{L}_{i=1} f_{i}
\sum^{M-1}_{\alpha=1} \sum^{M-1}_{\beta=1} \frac{\partial^{2} }
{\partial x_{i \alpha} \partial x_{i \beta}} 
\left[ {\cal D}_{\alpha \beta}(\underline{x_i}) P \right] \,,
\label{diffusion_M_Mut_L}
\end{eqnarray}
where ${\cal A}_{\alpha} (\underline{x})$ and ${\cal D}_{\alpha \beta}
(\underline{x})$ are given by Eqs.~(\ref{A_M_1}) and (\ref{D_M_1})
respectively.

An equation of a very similar form to Eq.~(\ref{diffusion_M_Mut_L})
has been obtained in connection with the model of language evolution
previously discussed~\cite{Baxter06}. In that case the equation was
derived through purely mesoscopic reasoning---there were no
``islands'' of $N$ tokens, only fractions, $x$, of different variants
in a speaker's grammar. In this case in order to derive the
Fokker-Planck equation, the mutation and migration rates also had to
be scaled, now not by $N$, but by $\lambda$, the weight given to new
tokens. The generational time had also to be scaled, this time by
$\lambda^{-2}$. Thus the scalings required to derive the Fokker-Planck
equation are identical in both cases.

\section{Backward-time formulation: coalescent theory}
\label{coalescent}

In Section~\ref{ideal} we described the Wright-Fisher model as a
forward-time process, i.e., one specifies the probability of having a
certain number of alleles in a subsequent generation \emph{given} the
number present in the current generation.  It is perfectly legitimate
instead to ask for the probability that a certain number of
individuals in the current generation are all descended from a set of
individuals from the \emph{previous} generation.  As we describe in
this section, one can reconstruct the ancestry of the present-day
population given that it has evolved under the dynamics defined in
Section~\ref{ideal}.  We shall also show below that this can be used
to recover any desired property of the population going forward in
time from a known initial condition.  Thus these forward- and
backward-time approaches are entirely equivalent.

\subsection{Reconstructing the ancestry}

To describe the backward-time approach in detail, we return to the
ideal Wright-Fisher population of $N$ haploid individuals described in
Section~\ref{ideal}.  Each of these individuals must have a parent
from the previous generation; some of them may share a parent, in
which case there are individuals in the previous generation that did
not have any offspring.  Hence, as one looks back in time the number
of ancestors of the present-day population must decrease until a
single ancestor of the entire population is found.

Let us assume that at $t$ generations in the past (note that in this
section, increasing $t$ means going back further in time) there are
$n$ ancestors of the present-day population.  The probability that any
pair of these ancestors have the same parent is $1/N$.  This can be
understood from the fact that when going forward in time, a new
generation can be constructed by assigning to each of the $N$
offspring a parent chosen at random from one of the $N$ individuals in
the previous generation.  The probability that out of any pair of
offspring, the second receives the same parent as the first is then
clearly $1/N$ as claimed.  If the number of ancestors is much smaller
than the total population size, $n \ll N$, the probability that three
or more individuals are siblings, or that that there is more than one
pair of siblings among the subpopulation of $n$ ancestors, is of order
$1/N^2$ and can thus be neglected in the limit $N\to\infty$.  As one
steps back a single generation, therefore, the number of ancestors
decreases by one with probability
\begin{equation}
\label{pc}
p_c(n) = {n \choose 2} \frac{1}{N} \;,
\end{equation}
since there are ${n \choose 2}$ pairs that could find a common parent;
otherwise the number of ancestors stays the same.

\begin{figure}
\begin{center}
\includegraphics[scale=0.5]{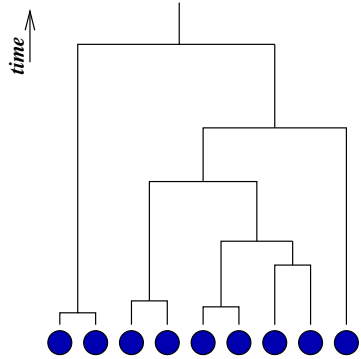}
\end{center}
\caption{\label{coalesce-ideal} Graphical representation of the
  coalescent process.  Each vertical line corresponds to a lineage
  leading to the final (present-day) population.  Two lineages merge
  whenever the ancestors they correspond to find a common parent.
  Since this coalescent rate is proportional to $n(n-1)$, where $n$ is
  the number of lineages remaining and is much smaller than the total
  population size $N$, the expected time between coalescence events
  considerably lengthens as one goes back in time.}
\end{figure}

It is customary to represent this decrease in the number of ancestors
in the form of a tree with the vertical direction representing time
(going upwards corresponds to going backward in time).  Each branch
represents an ancestor of the present day population and two branches
coalesce when the corresponding ancestors find a common parent---see
Fig.~\ref{coalesce-ideal}.  For this reason this backward-time process
was called ``the coalescent'' when formalised mathematically by
Kingman in the 1980's \cite{Kingman82}, although the underlying ideas
had previously been employed by population geneticists prior to this
point (some discussion of this earlier work can be found in
\cite{Tavare84}).

\begin{figure}
\begin{center}
\includegraphics[scale=0.5]{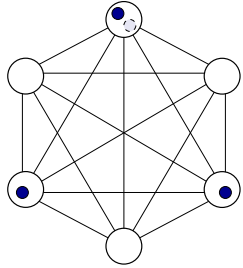}
\end{center}
\caption{\label{completegraph} Interpretation of the ancestral
  dynamics as a reaction-diffusion process on the complete graph.
  Each particle represents a lineage, and each particle hops to a
  neighbour, or stays where it is, with equal probability in each time
  step.  Should two particles ever coincide on the same site, they
  react immediately, leaving only a single particle.}
\end{figure}

Another way to picture this process, which may be more appealing for
physicists, is as a particle reaction system.  Let there be a lattice
(or, to be more technically correct, a graph) with $N$ sites (nodes),
$n$ of which are occupied by a particle.  In each timestep, each
particle either remains where it is, or hops with some probability to
one of its neighbours.  Should two particles ever occupy the same site
simultaneously, they immediately coalesce---see
Fig.~\ref{completegraph}.  One can verify that the
backward-time formulation of the ideal Wright-Fisher model
corresponds to this reaction-diffusion process on the complete graph
(i.e., a network in which each site is connected to every other---also
sometimes called a mean-field geometry).

From $p_c$ one can calculate statistics of the length of time that
elapses between a state with $n$ ancestors to a state with $m<n$
ancestors (as long as both $n$ and $m$ are much smaller than $N$).
This is because $p_c$ is constant until a coalescence occurs, and
hence the time spent in a state with $n$ ancestors has a geometric
distribution
\begin{equation}
\label{gd}
P_n(t) = [1-p_c(n)]^{t-1} p_c(n) \;.
\end{equation}
The mean of this distribution is thus $1/p_c(n)$, and since successive
coalescence events are uncorrelated, the time from $n$ to $m$
ancestors is
\begin{equation}
\label{Tnm}
T(n \to m) = \sum_{\ell = m+1}^{n} \frac{1}{p_c(l)} = 2 N
\sum_{\ell = m+1}^{n} \frac{1}{\ell(\ell-1)} = 2 N \left( \frac{1}{m}
- \frac{1}{n} \right) \;.
\end{equation}
This result shows that the ancestry of a present-day population is
dominated by an era in which the number of ancestors is small: the
mean time in which there are two ancestors $T(2 \to 1)$ is more than
half the time $T(m \to 1)$ to a single common ancestor from any number
$m \ll N$ ancestors.  In fact, it can further be shown that if one
starts with a sample comprising the entire population $m=N$, in the
limit of an infinite population $N\to\infty$, after any finite rescaled
time $\tau = t/N$, the number of ancestors remaining is finite (i.e.,
a vanishing fraction of the initial population) \cite{Griffiths84}.
This is significant because it means that the condition for only
pairwise coalescences with nonzero probability, i.e., that $n \ll N$,
is satisfied at any finite rescaled time $\tau$ in the past.

This backward-time description of Wright-Fisher population dynamics
has a number of practical benefits.  Firstly, the data that
geneticists have to hand comes from some present-day sample of
individuals, and thus it is sensible to condition the outcome of the
evolutionary dynamics on this known outcome.  Indeed, geneticists are
often interested in the information obtained by reconstructing the
ancestry of sampled genes, and a number of sophisticated methods based
on ensembles of ancestral trees have been developed with this aim in
mind \cite{Holder03}.  Secondly, it is very much more efficient to
simulate the backward-time process: to construct a genealogical tree
for $n$ present-day individuals with the desired rate involves
generating $n-1$ waiting times from the appropriate geometric
distributions (\ref{gd}).  Furthermore, having constructed this tree,
one can superimpose various of models of mutation (such as that
described above) since selectively neutral mutations (i.e., those that
do not cause individuals carrying a particular allele to have more
offspring than others) by definition do not affect the population
dynamics.  Finally, as we will discuss in Section~\ref{nonideal}
below, the backward-time method generalises more readily to
non-ideally mating populations (although one is restricted to
selectively neutral alleles).

\subsection{Relationship between forward- and backward time dynamics}

\begin{figure}
\begin{center}
\includegraphics[scale=0.5]{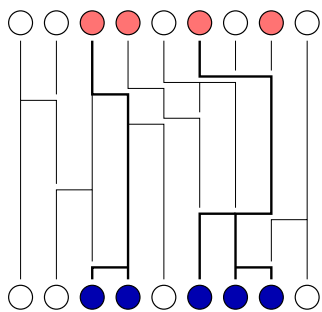}
\end{center}
\caption{\label{twosets} Lines of ascent/descent connecting two sets
  of individuals, one in the initial condition and one in the final
  condition.  Note here that shading does not refer to the allelic
  state of any individual, rather whether it is considered to be
  within or outside the set of interest at the early or late time.
  The heavy lines show the lines of descent from the early-time set
  that contribute to the late-time set.}
\end{figure}

Clearly, it should be possible to use the backward-time formulation
to calculate properties of the forward-time evolution from a fixed
initial condition.  The way to do this is to note that the process of
assigning mutant alleles to individuals in the initial condition is an
example of a (one-off) mutation process that has no effect on the
statistics of the line of descent.  Consider now Fig.~\ref{twosets}.
It shows the lines of descent from some initial condition: we can take
the individuals shown shaded at the earlier time (top) to be mutants,
and the remainder to carry the wild-type allele.  At the later time,
we can select some other set of individuals (e.g., that formed by the
dark-shaded individuals) and ask for the probability that all are
mutants.

Clearly for them all to be mutants, they must all be descended from
the mutants in the initial condition (if no mutation events take place
in the interim).  However, there may also be other mutants present at
the later time outside the set of interest (there are two such mutants
in Fig.~\ref{twosets}).  Therefore, to construct the probability
$R(a,d;t)$ that all $d$ sampled individuals at late time are descended
from one, more or all of the $a$ mutants at the initial time from
$P(b|a;t)$, the probability that there are \emph{precisely} $b$
mutants $t$ generations after the initial condition, one needs to sum
the latter, weighted by the probability that the chosen set falls
within the late-time mutant set.  Given that there are $b$ mutants in
the population, there are ${b \choose d}$ ways to choose $d$
individuals in such a way that all are mutants.  Since there are in
total ${N \choose d}$ possible ways in which $d$ individuals could
have been chosen, we have that
\begin{equation}
R(a,d;t) = \sum_b \frac{ {b \choose d} }{ {N \choose d} } P(b|a;t) \;.
\end{equation}

We can ask the same question using the backward-time language.  Let
$Q(c|d;t)$ be the probability that the late-time chosen set of $d$
individuals has coalesced into $c$ ancestors $t$ generations
previously.  For all $d$ late-time individuals all to be mutants, their
$c$ ancestors must all be drawn from the initial mutant set.  Using
the same kind of counting argument as previously one finds that
\begin{equation}
R(a,d;t) = \sum_c \frac{ {a \choose c} }{ {N \choose c} } Q(c|d;t) \;.
\end{equation}
Equating these two expressions gives an implicit duality relation
between the forward- and backward-time distributions $P$ and $Q$; this
can be found for example in \cite{Mohle01}.  Using the identity
\begin{equation}
\label{id}
\sum_{d=b}^{b'} (-1)^{b+d} \frac{{ N \choose b}{N-b \choose d-b} {b' \choose
d}}{{N \choose d}} = (-1)^b {b' \choose b} \sum_{d=b}^{b'} (-1)^d {b'-b
\choose d-b} = \delta_{b\,b'} \;,
\end{equation}
one can write this duality relation to give $P$ explicitly in terms of
$Q$ \cite{Blythe07}.  One finds
\begin{equation}
\label{PofQ}
P(b|a;t) = 
\sum_{c=1}^{a} \sum_{d=b}^{N}  (-1)^{b+d} \frac{{N \choose b}
  {N - b \choose d - b} {a \choose c}}{ { N \choose c}
  } Q(c|d; t) \;.
\end{equation}
Since it is usually the case that backward-time quantities are more
easily obtained than their forward-time counterparts (e.g., by Monte
Carlo sampling of the ancestral dynamics, or the equivalent
reaction-diffusion process), this formula provides one means for
efficient computation of the evolution from a fixed initial condition.
For example, if one wants to know the probability that a mutant allele
has become fixed by time $t$, one puts $b=N$ in the foregoing to find
\begin{equation}
P(N|a;t) = \sum_{c=1}^{a} \frac{{a \choose c}}{ { N \choose c} } Q(c|N;t) \;.
\end{equation}
Actually, in this case $Q(c|N;t)$ is known analytically
\cite{Griffiths84,Tavare84}, as indeed is $P(N|a;t)$, at least in the
limit $N\to\infty$ and after rescaling allele frequency $x_0=a/N$ and
time $\tau=t/N$ \cite{Crow70}.  Nevertheless, this approach has recently
found utility in applications to fixation times subdivided populations
in which neither of these distributions can be calculated exactly (see
\cite{Blythe07} and Section~\ref{nonideal} below).

\subsection{Scaling with population size in the coalescent
  formulation}

In Section~\ref{master} we saw that the characteristic timescale of
evolution grew linearly with population size for the Wright-Fisher
model, and quadratically in the Moran model.  One sees these same
scaling relations in the coalescent formulation.  Recall that above we
argued that the probability that two lineages coalesce in one
generation undergoing Wright-Fisher evolution is $p_c(2) =
\frac{1}{N}$.  Under the Moran model, the probability that a pair of
lineages coalesces is equal to the probability that one of them was
chosen to be the parent individual, and the other the offspring.  This
is $p_c(2)=\frac{2}{N^2}$ or $p_c(2) = \frac{2}{N(N-1)}$ if the
sampling is done with or without replacement, respectively.  The
factor of $2$ appears here because there are two ways in which a
(parent, offspring) permutation can be assigned to a pair of
individuals.

In both cases, the total rate of coalescence if there are $n$ lineages
present is $p_c(n) = {n \choose 2} p_c(2)$ as long as $n\ll N$.
Therefore in the limit $N\to\infty$ the ancestries of both
Wright-Fisher and Moran populations have the same dynamics as long as
time as rescaled as $\tau=t/N$ in the former and $\tau=2t/N^2$ in the
latter.  This is precisely the scaling that was previously used to
obtain the same Fokker-Planck equation for the two models.

In fact, very many neutral population dynamical processes have the
same ancestral dynamics as the Wright-Fisher and Moran models in the
limit $N\to\infty$ under the appropriate rescaling of time.  For
example, if one has a model in which the distribution of offspring
number for each individual is invariant under exchanges of the
individuals, and one can further define a timescale on which the
probability of ternary coalescences is vanishingly small compared to
binary coalescences, then (subject to other rather formal
requirements), the ancestral dynamics converges in the limit
$N\to\infty$ to that of the Wright-Fisher process (see, e.g.,
\cite{Mohle00} for the full mathematical details).  Population
geneticists refer to this property as the \emph{robustness} of the
coalescent; physicists will recognise this lack of sensitivity of the
late-time behaviour to details of the dynamics in large populations as
a kind of universality.

Furthermore, one can also understand the need to rescale migration and
mutation rates with the population size $N$ from the backward-time
formulation.  Consider the Moran model with two alleles as defined in
Section~\ref{WF_M_mut}.  The probability that a particular ancestor
was chosen as the offspring in the previous generation is $1/N$, and
further that the allele mutated after the parent was sampled is either
$u$ or $v$ (depending on the allele in question).  For coalescence and
mutation probabilities to be of the same order in $N$, we therefore
require $u$ and $v$ both to be of order $1/N$.

\section{Non-ideal populations}
\label{nonideal}

We have thus far mostly restricted our discussion to ideal
populations, i.e., those in which the individual(s) displaced through
the production of new offspring are chosen from a uniform
distribution.  We now turn to the contrasting case of non-ideal
populations, where the individuals to displace are chosen
non-uniformly and investigate the various ways in which population
subdivision has been handled by population geneticists.

Non-ideal populations are most easily handled within the backward-time
formulation, and therefore it is most convenient to describe the
migration events in terms of the set of probabilities $\mu_{ij}$ that
any given individual randomly sampled from subpopulation (or deme) $i$
is the offspring of a parent individual from the previous generation
in deme $j$.  For the case $i\ne j$, this is given simply as the joint
probability for a parent to be in deme $j$ and an offspring in deme
$i$, which in our earlier notation is
\begin{equation}
\mu_{ij} = f_j g_{ij} \;.
\end{equation}
On the other hand, the probability that an individual in deme $i$ is
not an immigrant is obtained from the fact that $\sum_{j} \mu_{ij} =
1$, viz,
\begin{equation}
\mu_{ii} = 1 - \sum_{j \ne i} f_j g_{ij} \;.
\end{equation}

It is worth at this stage to explore these migration parameters a
little further.  First, if subpopulation $i$ comprises $N_i$
individuals, the expected number of immigrants per generation is
\begin{equation}
N_i^{\rm{(in)}} = N_i \sum_{j \ne i} \mu_{ij} \;.
\end{equation}
Meanwhile, the expected number of emigrants from deme $i$ (defined as
the number of offspring of individuals that end up in a deme $j\ne i$)
is
\begin{equation}
N_i^{\rm{(out)}} = \sum_{j \ne u} N_j \mu_{ji} \;.
\end{equation}
Migration is called \emph{conservative} if the expected number of
individuals entering and leaving every deme per generation is the
same, i.e., if $N_i^{\rm{(in)}} = N_i^{\rm{(out)}} \; \forall\,i$, or
equivalently, if
\begin{equation}
\label{cons}
\sum_{j \ne i} N_i \mu_{ij} = \sum_{j\ne i} N_j \mu_{ji}
\quad\forall\,i \;.
\end{equation}
Note that this corresponds with the definition of conservative
migration previously given (\ref{consG}) when all deme sizes
are equal, $N_i=N$.  Now, let us briefly consider the common ancestor
of a subdivided population.  The single lineage that remains as
$t\to\infty$ will hop from between subpopulations at a rate $\mu_{ij}$
from deme $i$ to deme $j$.  Therefore, the stationary distribution of
the common ancestor (assumed unique) is given by the solution of the
set of equations
\begin{equation}
\label{migbal}
\sum_{j \ne i} Q_j^\ast \mu_{ji} = \sum_{i \ne j} Q_i^\ast \mu_{ij} \;.
\end{equation}
We see that if migration is conservative, i.e., if (\ref{cons}) holds,
the probability that the common ancestor is in subpopulation $i$ in
the steady state is proportional to its size,
\begin{equation}
Q_i^\ast = \frac{N_i}{L \bar{N}}
\end{equation}
where $L$ is the number of subpopulations and $\bar{N} = \frac{1}{L}
\sum_{i=1}^{L} N_i$ is their mean size.

Clearly, one can construct models of varying complexity and intricacy
by suitable choices of the migration probabilities $\mu_{ij}$ and
subpopulation sizes $N_i$, although some cases are still excluded from
this description.  Consider the following classical examples.

\paragraph{Symmetric island model} The simplest model of subdivision
has $L$ equal-sized subpopulations (islands) of equal size $N_i=N$
from which a uniform fraction $\mu$ of the individuals on each island
are replaced by offspring chosen at random from the other islands:
$\mu_{ij} = \mu/(L-1)$ for all pairs $i \ne j$.  The version of this
model that has an infinite number of islands was introduced by Wright
\cite{Wright31,Wright43}; the finite island model is often attributed
to Maruyama \cite{Maruyama70a} and Latter \cite{Latter73}.

\paragraph{Randomly-mating diploid population with two sexes} Recall
that a diploid population is one that carries two instances of each
gene which may or may not be the same allele.  A diploid population of
$L$ individuals can be thought of as a set of $L$ subpopulations, each
of size $2$.  One gene from in each subpopulation is sampled randomly
from the $L_m$ males in the populations; the other from the
$L_f=L-L_m$ females.  To represent this model via the migration
parameters $\mu_{ij}$, however, it is necessary to have $2L$
subpopulations each of size $1$.  We then take $\mu_{ij} = \frac{1}{2
L_m}$ if subpopulation $i$ is a gene that is inherited from a male and
$j$ is one of the $2L_m$ genes carried by a male in the previous
generation; and $\mu_{ij} = \frac{1}{2 L_f}$ is the corresponding
quantity for females.

\paragraph{Randomly-mating asexual diploid population with
hermaphrodism} As an alternative to sexual reproduction, the two genes
carried by a diploid individual could be inherited from the same
parent (through a process called \emph{selfing}) with probability $S$,
or from two different parents with probability $1-S$ (by
\emph{outcrossing}).  This model cannot be reproduced using the
$\mu_{ij}$ parameters because the parental distribution is not
independent for the two genes carried by a single individual.

\bigskip

It is also possible for populations to be subdivided by age group, for
example, if only individuals within some age class are fertile.  Given
that considerable complexity is possible within even the simple
general model of population subdivision outlined here, it is
unsurprising that geneticists have sought a small number of parameters
with which to characterise and differentiate the various specific
cases that can be constructed.  Two quantities that appear prominently
throughout the population genetics literature are \emph{inbreeding
coefficients} and \emph{effective population size}.  Both of these
concepts are commonly said to have been introduced by Wright
\cite{Wright51}.

\subsection{Inbreeding coefficients, fixation indices and $F$-statistics}
\label{inbreeding_coeff}

Geneticists are often most interested in diploid organisms, and there
an important question is whether the two instances of a gene at a
specified locus are the same or not.  In the absence of mutation, the
two genes are the same only if they share a common ancestor (a
situation described as being \emph{identical by descent}).  Since this
is the case if some \emph{inbreeding} has occurred, the probability
$F$ that two genes are identical by descent is sometimes called an
\emph{inbreeding coefficient}.

Given a sample of some individuals taken from a population, $F$ can be
estimated by examining specific \emph{genetic markers}.  For example,
one can look for different enzymes that can catalyse a particular
chemical reaction, or look directly at DNA sequences (see e.g.,
\cite{Neigel97} for a description of various methods in the context of
subdivided populations).  Of particular interest is the way in which
population structure manifests itself in observed identity
probabilities: typically, one expects there to be a greater chance for
genes sampled from a subpopulation to be identical, than those from
the wider (meta)population.

A quantity that encodes this information and that has become
ubiquitous in analyses of population structure is denoted $F_{ST}$,
variously called an inbreeding coefficient, fixation index or
$F$-statistic, and was introduced by Wright \cite{Wright51}.  A number
of subtly different, sometimes ambiguous, definitions of this quantity
can be found dotted around the literature---one senses a slight air of
frustration arising from this state of affairs in the introduction of
\cite{Nagylaki98} for example.  Following the approach of
\cite{WilkinsonHerbots98} (and others), we avoid getting too heavily
into the distinctions between different expressions for $F_{ST}$, and
simply adopt a particular definition, namely that provided by Nei
\cite{Nei73} which reads
\begin{equation}
\label{FST}
F_{ST} = \frac{F_0 - \bar{F}}{1 - \bar{F}} \;.
\end{equation}
Here $F_0$ is the probability that two genes drawn at random from the
same subpopulation are identical by descent, and $\bar{F}$ is the
corresponding quantity for a pair of genes sampled randomly from the
entire population.  Nevertheless this definition leaves open the
question of whether these samples should be weighted according to the
subpopulation sizes or not; in practice, this depends on how $F_{ST}$
has been determined from an actual sample at hand.

Despite its ubiquity and utility, there are various aspects of this
quantity that are confusing at a first encounter.  First there is the
(rarely elucidated) notation: $S$ and $T$ simply stand for
subpopulation and total population respectively.  With reference to
diploid populations, one can also define analogous quantities $F_{IS}$
and $F_{IT}$ which measure how much more (or less) closely related
the pair of genes contained within an individual are than a pair
sampled from the population at large.

Next, one often finds a single expression of $F_{ST}$ quoted for a
given model of subpopulation, even though identity probabilities in
general change with time.  Wright's infinite island model (defined
above) provides one example of a case where $F_{ST}$ is time
independent.  Since the total population size is infinite, the number
of individuals across the whole population that carry a particular
allele does not change with time.  However, if one takes a subsample
of this population of size $N$, there will be variation in the number
of alleles of a particular type, and so one finds $F_0 \ne \bar{F}$
which leads to a nontrivial $F_{ST}$.  Wright \cite{Wright43} obtained
an expression for $F_{ST}$ by taking the case of two alleles with a
fraction $x_i$ of the individuals in subpopulation $i$ carrying one of
them (say, the wild type).  In such a case, one can show that
\begin{eqnarray}
F_0 &=& \frac{1}{L} \sum_i \left[ x_i^2 + (1-x_i)^2 \right] = 2
\overline{x^2} - 2 \bar{x} + 1 \\
\bar{F} &=& \frac{1}{L^2} \sum_{i,j} \left[ x_i x_j + (1-x_i)(1-x_j)  \right] =
  2 \bar{x}^2 - 2 \bar{x} + 1
\end{eqnarray}
where $\overline{f(x)} = \frac{1}{L} \sum_i f(x_i)$.
Hence,
\begin{equation}
\label{FSTvar}
F_{ST} = \frac{\overline{x^2} - \bar{x}^2}{\bar{x}(1-\bar{x})} \;.
\end{equation}
We remark that in the limit of infinite subpopulations (where there is
no distinction between sampling with and without replacement), one has
$0 \le F_{ST} \le 1$, the lower and upper bounds relating to extremes
of homogeneous and heterogeneous spatial distributions of alleles
respectively.  These bounds one can see by noting first that the
numerator of (\ref{FSTvar}) is positive, since it is a variance, and
second that since all $0 \le x_i \le 1$ it follows that $0 \le \bar{x}
\le 1$ and further that $\overline{x^2} \le \bar{x}$.  When
subpopulations are finite, it is possible for $F_{ST}$ to be slightly
negative as a consequence of the effects of sampling without
replacement: having removed one gene from a subpopulation, the
probability of finding an identical gene in another subpopulation
could be of order $1/N$ \emph{more} likely than in the same
subpopulation.  Therefore, if one uses (\ref{FSTvar}) as a general
definition of $F_{ST}$ rather than (\ref{FST}), as is done for example
in \cite{Nagylaki98}, the results obtained will not necessarily agree
in situations where subpopulations are finite.

Using (\ref{FSTvar}), Wright \cite{Wright43} showed by an explicit
calculation of the variance in the allele frequencies that for the
island model
\begin{equation}
\label{WIMFST}
F_{ST} = \frac{(1-\mu)^2}{N - (N-1)(1-\mu)^2} \;.
\end{equation}
A quick way to obtain this result is to focus on a single
subpopulation and assume that any immigrants that arrive from the
wider population do not have a common ancestor (and are hence
unrelated).  This assumption can be justified by switching to the
coalescent picture outlined in Section~\ref{coalescent}, suitably
extended to cater for population subdivision.  If two lineages are
situated within a single subpopulation, they coalesce at a rate $1/N$,
whereas two lineages in different subpopulations will coalesce at a
rate proportional to $\mu/L$.  If the subpopulation size $N$ is held
fixed whilst $L\to\infty$, a separation of timescales occurs and the
probability that two lineages in different subpopulations coalesce
effectively vanishes.  Hence, $\bar{F}$ is effectively zero (since the
chance of two randomly-chosen individuals are from the same
subpopulation is also of order $1/L$), and $F_{ST}=F_0$.  Using now
the fact that $F_0$ does not change with time, one can establish that
\begin{equation}
F_0 = (1-\mu)^2 \left[ \frac{1}{N} + \left( 1-\frac{1}{N} \right) F_0
  \right]
\end{equation}
since the right-hand side of this equation gives the value of $F_0$
after one generation of the Wright-Fisher process (as defined in
Section~\ref{ideal}).  This comes about because the probability that
the members of any pair are both descended from individuals from the
subpopulation of interest is $(1-\mu)^2$; the probability that both
are offspring of the same parent is $1/N$; and if they are offspring
of different parents, those parents were themselves related with
probability $F_0$.  Rearranging this equation yields (\ref{WIMFST}).
Note that within this picture, $F_{ST}$ can be interpreted as the
probability that two lineages in the same subpopulation coalesce
before one of them migrates away (looking backward in time).

It is worth looking more closely at how $F_{ST}$ behaves in the
various limits that have previously been established.  First, if the
migration probability $\mu$ remains finite as the subpopulation size
$N\to\infty$, $F_{ST}\to 0$ which indicates an absence of spatial
structure in gene frequencies---precisely what one would expect in the
strong migration limit.  On the other hand, if the combination $N\mu$
is finite as $N\to\infty$, one has
\begin{equation}
F_{ST} \approx \frac{1}{1+2\mu N}
\end{equation}
to leading order in $1/N$.  As a consequence of this relationship, a
value of $F_{ST}$ estimated by sampling from a population is often
used as a means to obtain the mean \emph{number} of migrants $\mu N$
arriving in a subpopulation in each generation, this being considered
an easier task practically than tracking the movement of individuals
between subpopulations.  We also remark that (\ref{WIMFST}) is not
restricted to the special case where the migration rate between every
pair of islands is the same, since it has only been assumed that $F_0$
is stationary, and that all immigrants into the subpopulation of
interest are unrelated.  In practice this means one requires an
infinite total population that is at equilibrium, or has stationary
mean allele frequencies as a consequence of conservative migration.
Nevertheless, these are still quite restrictive assumptions and it
turns out that estimating $\mu N$ in this indirect way is fraught with
difficulties \cite{Whitlock99}.

There is therefore much interest in exploring how $F_{ST}$ behaves
under more general conditions, and in particular under models of
population subdivision in which migration is not conservative, or the
number of subpopulations is finite.  In these models, allele
frequencies vary with time, and typically therefore the focus is on
the value of $F_{ST}$ that is approached as equilibrium is reached.
In a model with a finite total population, the equilibrium state is
one of fixation, as previously discussed in Section~\ref{ideal}.
Hence all identity probabilities tend to unity.  However, the ratio
that appears in (\ref{FST}) approaches a nontrivial finite value, and
thus contains some information about population structure that can be
related to gene diversity.

A popular way to calculate this limiting value is to exploit a trick
due to Slatkin \cite{Slatkin91}.  Let us trace the ancestry of a pair
of genes sampled from the present day population.  From the discussion
of Section~\ref{coalescent} we know that eventually a common ancestor
of these genes are found; let us suppose that this happens at time $t$
with probability $c(t)$.  Let us also introduce a Poisson mutation
process that occurs with probability $\theta$ per generation.  Then,
the probability that the two genes are \textit{identical in state}
(i.e., share a common ancestor \emph{and} have not been subject to any
mutations since the time of coalescence) is
\begin{equation}
F(\theta) = \sum_{t=1}^{\infty} c(t) (1-\theta)^{2t} \;.
\end{equation}
In the limit $\theta\to0$, the concepts of identity by descent and
identity in state become equivalent, and so asymptotically,
\begin{equation}
\label{FSTT}
F_{ST} = \lim_{\theta\to0} \frac{F_0(\theta) - \bar{F}(\theta)}{1 -
  \bar{F}(\theta)} = 1 - \frac{T_0}{\bar{T}}
\end{equation}
where $T_0$ and $\bar{T}$ are the mean times for two genes randomly
sampled from within or between subpopulations respectively to find a
common ancestor.  In essence, therefore, the task is to calculate the
mean first passage time for a pair of random walkers which hop between
sites of a network at rates controlled by the migration parameters
$\mu_{ij}$.

This is achieved by solving the set of linear equations
\begin{equation}
\label{Teqs}
T_{ij} = 1 + \sum_{k \ell} \mu_{ik} \mu_{j\ell} T_{k \ell}\left(1 -
\frac{1}{N_k} \delta_{k,\ell}\right)
\end{equation}
where $T_{ij}$ is the mean time for two lineages, one in subpopulation
$i$ and one in subpopulation $j$ to coalesce.  This equation arises
because if one has two distinct lineages, one needs to wait at least
one generation for a change of state to occur.  The joint probability
that the lineages in $i$ and $j$ hop to $k$ and $\ell$ is
$\mu_{ik}\mu_{j\ell}$.  If they arrive in different subpopulations, we
must then further wait a mean time $T_{k \ell}$ for coalescence to
occur.  On the other hand, if they land in the same place $k$, there
is a probability $1/N_k$ for coalescence to occur immediately (i.e., a
mean waiting time of zero), otherwise we need to wait $T_{kk}$ on
average.  As with $F_{ST}$, the question arises as to how to form the
averages $T_0$ and $\bar{T}$ appearing in (\ref{FSTT}).  For example,
one could take $T_0 = T_{ii}$ for a specific subpopulation of
interest.  Alternatively, one could average over all subpopulations,
possibly weighted according to their size.  It is not uncommon to see
both size-weighted and unweighted versions of $\bar{T}$ in the
literature, presumably because the correct form depends on whether
subpopulation sizes are known experimentally or not (some further
discussion is given in \cite{WilkinsonHerbots98}).

For arbitrary migration rates (\ref{Teqs}) can be rather cumbersome to
solve, even for simple models of population subdivision.
Simplifications occur in the limit of slow migration, where $\mu_{ij}
= m_{ij}/\bar{N}$ where $\bar{N}$ is the mean subpopulation size and
is taken to infinity.  Recall from Section~\ref{ideal} that this is
the regime in which both drift and migration operate on the same
timescale in the limit of an infinite population size; meanwhile, in
the backward-time formulation, this scaling has migration events
occurring on the same timescale as coalescence (see, e.g.,
\cite{Nordborg03}).  In this limit, terms which have both lineages
hopping simultaneously do not enter into (\ref{Teqs}).  Furthermore,
it has been shown \cite{Strobeck87} that, if migration is
conservative, the mean time for coalescence of lineages located in the
same subpopulation $i$ is independent of $i$ and given by
$T_{ii}=L\bar{N}$.  This fact considerably simplifies calculation of
$F_{ST}$ for conservative models.  For example, Slatkin
\cite{Slatkin91} showed that for the symmetric island model with a
finite number $L$ of islands undergoing a slow migration process,
\begin{equation}
F_{ST} = \frac{1}{1+2Nm \left(\frac{L}{L-1}\right)^2} \approx
\frac{1}{1 + 2Nm} \left[1 + \frac{2}{1+\frac{1}{2Nm}} \frac{1}{L} +
  O\left(\frac{1}{L^2}\right) \right]
\end{equation}
which agrees with the result previously obtained by other methods
(such as that used by \cite{Latter73}; see also
\cite{WilkinsonHerbots98}).  We observe the the finite-size
corrections to the infinite island result, (\ref{WIMFST}), are of
order $1/L$ which perhaps sheds some light on the ubiquity of the
expression (\ref{WIMFST}) in analyses of population structure.

This procedure for calculating the fixation index $F_{ST}$ has been
followed for a range of models of population subdivision, such as a
hierarchical island model \cite{Slatkin91b} (which has one finite
island model nested inside another), a model with two subpopulations
of different sizes \cite{WilkinsonHerbots98} and a continental-island
model, where migration does not take place directly between the
islands but only via a central continent \cite{WilkinsonHerbots98}.
These models are tractable because of some symmetry or regularity in
the migration structure that can be exploited.  It is also possible to
make progress in cases where the number of subpopulations tend to
infinity \cite{WA01,WT04,WL06}.  However, we will not discuss these
models further here.

\subsection{Effective population size}
\label{effective_pop}

As we described in the previous section, the motivation behind
studying $F_{ST}$ is that it is a statistic of a population that can
be estimated by examining appropriate data.  A second characteristic
of biological populations that has also received considerable
attention is its \emph{effective size}, $N_e$.  Like $F_{ST}$, this
notion can also be attributed to Wright \cite{Wright31}, although it
is also understood that Crow and coworkers also made significant
contributions to its development (see, e.g., the classic text
\cite{Crow70}).  The basic idea here is that an ideal population is
entirely characterised by the number of individuals $N$ that are
randomly mating.  In a real population, one suspects that fluctuations
(like those associated with genetic drift) might be due to some
special ``breeding'' individuals that form a subset of the full
population.

There are as many ways to define $N_e$ as there are properties of a
population that one might be interested in; furthermore, these
different definitions will not generally give the same value for a
given model of population subdivision.  Nevertheless, for the concept
to be useful, one hopes that the effective size determined by
examining one quantity is indicative of how the population as a whole
will behave.  It is not appropriate here to get too involved in all
the various definitions and their pros and cons; instead we will
highlight two particularly prominent definitions and give a flavour of
how they depend on certain aspects of the population structure.  For
more extension discussions we refer the reader to
\cite{Wang99,Rousset04}.

One of the most natural definitions of effective size arises from the
forward-time prescriptions of the population dynamics given in
Section~\ref{ideal}.  Recall that under the Wright-Fisher dynamics, if
the allele frequency in one generation is known to be $x'$, in the next
it is a random variable $x$ that has variance $x'(1-x')/N$.  This
connection between the variance of an allele frequency and population
size can be exploited to define an effective population size in a more
general context.  Let $\bar{x}'$ be some measure of an allele frequency
across a subdivided population, e.g., the arithmetic mean
$\bar{x}'=\frac{1}{L} \sum_{i=1}^{L} x_i'$, and $\Var(\bar{x})$ the
variance in this quantity after one generation of the underlying
population dynamics.  An effective size $N_e$ of this subdivided
population can then be given as
\begin{equation}
N_e = \frac{\bar{x}'(1-\bar{x}')}{\Var(\bar{x})} \;.
\end{equation}
To distinguish it from other effective sizes than can be defined, this
one is called the \emph{variance effective size}.  Note that in this
expression the variance is over the distribution of allele frequencies
$\bar{x}$ generated by the population dynamics, and \emph{not} the
variance over subpopulations appearing in (\ref{FSTvar}).  A drawback
of this definition of $N_e$ is that it is likely to depend on the
frequency $\bar{x}$, and possibly other statistical quantities---such
as the variance of allele frequencies over subpopulations.  Therefore,
if one seeks a description of the dynamics that is closed in a small
set of random variables, it will typically have to be an approximate
description (see, e.g., \cite{Roze03,Rousset03,Cherry03}).

A definition of effective population size that avoids these
difficulties makes use of the fact that when tracing lineages of an
ideally mating population backward in time, pairs coalesce at a rate
$1/N$.  The mean rate of coalescence between pairs of lineages in a
subdivided population can then be taken to define the reciprocal of an
effective population size.  As with the fixation index $F_{ST}$, this
rate will typically vary with time; in those cases, one can adopt the
coalescence rate that is approached as one looks infinitely far back
in time.  The effective size that results is then often referred to as
an \emph{asymptotic effective size}.  In fact, the same limit has been
shown to be approached by a range of different definitions of
effective size \cite{Whitlock97,Rousset04}.  This fact can traced to
all definitions being governed asymptotically by the decay rate of the
longest-lived nonstationary eigenstate of the underlying population
dynamics.  Furthermore, this feature makes the asymptotic effective
population size an appealing quantity to devote some effort to
calculating.

As with $F_{ST}$, there are a few cases where the mean coalescence rate
does not change with time and can be easily be calculated exactly.
The first two such cases were introduced at the start of this section.

\begin{figure}
\begin{center}
\includegraphics[scale=0.66]{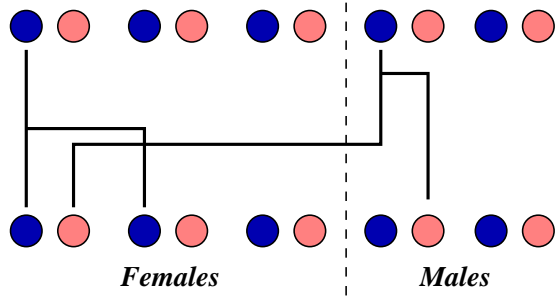}
\end{center}
\caption{\label{coalesce-mf} Sexually reproducing diploid population.
Each individual carries one gene inherited from a female (shown solid)
and one from a female (shown shaded).  Two ways in which two distinct
genes may coalesce are shown.  In both cases, the two offspring must
have been inherited from an individual of the same sex.}
\end{figure}

\paragraph{Randomly-mating diploid population with two sexes}  If
there are two distinct lineages in a diploid population, they must
both be inherited from the same parent (and furthermore be the same
gene from that parent) if they are to coalesce---see
Fig.~\ref{coalesce-mf}.  Taking two genes from the population at
random (with replacement for simplicity), the probability that both
were inherited from a male, or both from a female, is $\frac{1}{4}$.
Within these classes, the same gene is chosen with probability
$\frac{1}{2L_m}$ and $\frac{1}{2L_f}$ for male and female parents
respectively.  Hence one finds a total coalescence probability per
generation of
\begin{equation}
\frac{1}{4} \left( \frac{1}{2L_m} + \frac{1}{2L_f} \right) = \frac{L_m
  + L_f}{8 L_m L_f} = \frac{1}{N_e} \;,
\end{equation}
where we have identified the coalescence rate with an inverse
population size as described above.  This result was first given by
Wright \cite{Wright31}.  Notice that $N_e \le 2 (L_m+L_f)$, that is,
if there is any imbalance in the number of males and females, the
effective number of gene instances in the population is less than
their actual number.

\begin{figure}
\begin{center}
\includegraphics[scale=0.66]{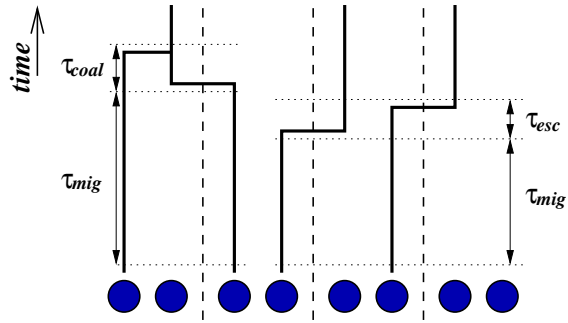}
\end{center}
\caption{\label{coalesce-herm} Separation of timescales in a
hermaphrodite diploid population.  The typical timescale $\tau_{\rm
mig}$ for two lineages to migrate into the same individual is very much
longer than the typical time to coalescence $\tau_{\rm coal}$ due to
selfing, or to escape to different individuals $\tau_{\rm esc}$ due to
outcrossing.}
\end{figure}

\paragraph{Randomly-mating asexual diploid population with
  hermaphrodism} In this model, the effective size can be calculated
as a consequence of a separation of timescales. A classical
calculation can be found in \cite{Pollak87}; here we follow the
coalescent-based approach introduced in \cite{Nordborg97}, also
outlined in \cite{Nordborg03}.  As we have already remarked, this
model is subtly different to that described in Section~\ref{ideal}, in
that the parents of two lineages are not chosen independently.
Instead, should two lineages find themselves within the same
individual there is a selfing probability $S$ that both are descended
from genes contained within a single parent individual, and $1-S$ that
they are each descended from different individuals.  At the time of
arrival into a single individual, these two lineages have a probability
$1/2$ of immediately coalescing; otherwise, going back a further
generation there is a probability $S/2$ that they are both still in
the same individual, uncoalesced, or $S/2$ of coalescing, or $1-S$ for
the two lineages to return to a state at which they are located in
different individuals.  Using the properties of the geometric
distribution, it can be shown that two lineages arriving at the same
individual have a total coalescence probability of $1/(2-S)$, and the
mean time to do so is $S/(2-S)$; on the other hand, escape occurs with
probability $(1-S)/(2-S)$ and takes on average $2/(2-S)$ generations
\cite{Nordborg97,Nordborg03}.  Meanwhile, when two lineages are in
different individuals, the mean time for them to arrive in the same
individual is $L$, the number of individuals (since the arrival
probability per generation is $1/L$).  Thus, once $L$ is taken
sufficiently large at any fixed $S$, the rate of coalescence upon
arrival is very much faster than the rate of arrival---see
Fig.~\ref{coalesce-herm}.  Then, one can approximate the coalescence
rate as the product of the arrival rate and the subsequent probability
of coalescence, i.e.,
\begin{equation}
\frac{1}{N_e} = \frac{1}{L(2-S)} \;.
\end{equation}
Again, the effective (haploid) population size $N_e = L(2-S)$ is less
than twice the actual population size, unless there is no selfing
($S=0$), when the effective and actual sizes are equal.

\paragraph{Population subdivision in the strong migration limit}
The coupling of a large ratio of coalescence to migration rates and
small subpopulation sizes is crucial in admitting the computation of
effective size of a hermaphrodite population.  In the opposite limit,
where migration is very fast compared to coalescence and subpopulation
sizes are large, a simple, expressions valid for general models of
population subdivision can also be obtained.  The key is to realise
that between coalescence events, all lineages are independently
distributed according to the stationary distribution generated by the
backward-time migration dynamics.  That is, we assume that the
probability for two lineages to be present in subpopulation $i$ is
$(Q_i^\ast)^2$ where $Q_i^\ast$ is the stationary distribution for the
single common ancestor of the whole population given by the migration
balance equations (\ref{migbal}).  Given that the rate of coalescence
for two lineages in subpopulation $i$ is $1/N_i$ it then follows that
under this approximation that the mean coalescence rate is
\begin{equation}
\frac{1}{N_e} = \sum_{i=1}^{L} \frac{(Q_i^\ast)^2}{N_i} \;.
\end{equation}
This formula has been shown to exact (to leading order in $1/N$) if
the migration probabilities $\mu_{ij}$ obey 
\begin{equation}
\lim_{\bar{N} \to \infty} \bar{N} \mu_{ij} = \infty
\end{equation}
where $\bar{N}$ is the mean subpopulation size \cite{Nagylaki80}.
This equation defines a strong migration limit, within which it is
seen that the effective population size is always less than or equal
to the total population size $L\bar{N}$.  Equality is reached only if
the stationary probabilities $Q_i^\ast$ are in proportion to the
subpopulation size, i.e., if $Q_i^\ast = N_i/(L \bar{N})$.  If the
stationary distribution is unique, this is the case only if migration
is conservative.  Since ancestral lineages are well-mixed, one does
not see any variation between subpopulations, and so the approximation
of the entire population as an ideal population with a possibly
reduced effective size would appear to be a good one.

\bigskip

For more general models of subdivision, where migration probabilities
are typically of order $1/\bar{N}$, fewer exact results are available.
Nevertheless, a formula for the effective population size in terms of
the asymptotic changes in identity probabilities has been given
\cite{Rousset04} as
\begin{equation}
\label{Ne}
\frac{1}{N_e} = \lim_{t\to\infty} \frac{\sum_{i=1}^{L} \frac{1}{N_i}
  (Q_i^\ast)^2 (1 - F_{ii}(t))}{ \sum_{i=1}^{L} \sum_{j=1}^{L}
  Q_i^\ast Q_j^\ast (1-F_{ij}(t))}
\end{equation}
where $F_{ij}(t)$ is the probability that two individuals, one sampled
from subpopulation $i$ and one from subpopulation $j$, are identical
by descent after $t$ generations, as defined in the previous
subsection.

This formula takes a particularly simple form in models when migration
is conservative.  Then, as shown at the start of this section, the
probability $Q_i^\ast$ that the single remaining ancestor is in deme
$i$ in the steady state is proportional to $N_i$, the size of deme
$i$.  Then, reduces to
\begin{equation}
\frac{1}{N_e} = \frac{1}{LN} \frac{1-F_0}{1-\bar{F}}
\end{equation}
where $F_0$ and $\bar{F}$ here are \emph{deme-size-weighted} probabilities
of identity for pairs of genes sampled from within demes and from the
whole population respectively, that is
\begin{eqnarray}
F_0 &=& \frac{1}{L} \sum_{i=1}^{L} \frac{N_i}{\bar{N}} F_{ii}\\
\bar{F} &=& \frac{1}{L^2} \sum_{i=1}^L \sum_{j=1}^{L} \frac{N_i}{\bar{N}}
\frac{N_j}{\bar{N}} F_{ij} \;.
\end{eqnarray}
Defining a size-weighted $F_{ST}$ through (\ref{FST}) and the
formul\ae\ for $F_0$ and $\bar{F}$ given here, we find that for
conservative models $N_e$ and $F_{ST}$ can be related through
\begin{equation}
N_e = \frac{L\bar{N}}{1 -  F_{ST}} \;.
\end{equation}

When the assumption of conservative migration is relaxed, one does not
find a simple relation between effective population size and
inbreeding coefficients.  Furthermore, it is not clear that a single
parameter (such as $N_e$ of $F_{ST}$) should satisfactorily
characterise all aspects of the evolutionary dynamics.  For example,
when migration is weak, it has been argued that the ancestral dynamics
can only be fully expressed in terms of takes ``structured
coalescent'' \cite{Notohara90} that takes into account the numbers of
lineages present in a particular subpopulation at finite times rather
than the weighted late-time average implied by (\ref{Ne})
\cite{Nordborg02,Sjodin05}.  In such cases, it is therefore necessary
to examine specific models.  For example, the variation of fixation
times in structured populations undergoing weak migration has been
investigated in an exact approached based on the duality relation
given in Section~\ref{coalescent} for an ideal population
\cite{Blythe07}.  This allows for a comparison with the fixation time
that would be seen in an ideal population with an effective size given
by (\ref{Ne}).  In a range of models considered, it was found that the
effective size gives an extremely good prediction for the fixation
time, except in a particular case where special properties of the
network structure admitted fixation to occur in a finite time in an
infinite population.  This finite-time effect has also been seen in
voter models on heterogeneous (scale-free) graphs
\cite{Sood05,Antal06}, which in fact is a special case of the general
model of population subdivision outlined here with all subpopulation
sizes $N_i=1$.

\section{Selection}
\label{selection}

So far we have considered only neutral models of evolution, that is,
those for which there is no preference for a particular allele.
Despite being apparently a reasonable model for some aspects of
genetic, ecological or linguistic behaviour (as we have previously
discussed) geneticists in particular have been interested in the fate
of alleles that are selected for or against.

The relationship between the genetic make-up of an individual and its
survival is of course very complicated.  However, one can explore the
effects of selection by simply introducing parameters that determine
how many offspring an individual carrying a particular allele (or
combination of alleles when diploid organisms are being considered)
has on average.  In this section we offer a small taste of some
evolutionary models that encompass selection.

Let us return to the model that has a randomly-mating haploid
population of $N$ individuals with two alleles, denoted $A$ and $B$.
In the neutral model, the parent of each individual in an offspring
generation is chosen uniformly from all possible parents.  When
selection is active, it is supposed that an individual with allele $A$
($B$) is chosen to be a parent with a weight $w_A$ ($w_B$).  That is,
if there are $n'$ $A$ alleles in the parent generation, each offspring
has a probability
\begin{equation*}
\frac{n' w_A}{n' w_A + (N-n') w_B}
\end{equation*}
of carrying allele $A$ (at least, if no mutations occur).  Since each
individual is assigned a parent independently, we have that the
probability for there to be $n$ $A$ alleles in the next generation is
\begin{eqnarray}
p_{n \,n'} &=& {N \choose n} \left( \frac{n' w_A}{n' w_A + (N-n') w_B}
\right)^n \left( \frac{(N-n') w_B}{n' w_A + (N-n') w_B}
\right)^{N-n} \\
&=& \frac{1}{(\bar{w}')^N} {N \choose n} \left( \left[
  \frac{n'}{N} \right] w_A \right)^n \left( \left[ 1 - \frac{n'}{N}
  \right] w_B \right)^{N-n} 
\label{Sel_Trans_Prob}
\end{eqnarray}
where in the second line we have simplified the notation by
introducing the mean \emph{fitness} of a population with $n'$ $A$ alleles
\begin{equation}
\bar{w}' = \frac{1}{N} \left( n' w_A + [N-n'] w_B \right) \;.
\end{equation}
Note that when the two alleles have equal fitness, i.e., $w_A=w_B$,
(\ref{Sel_Trans_Prob}) reduces to (\ref{WF_trans_Probs}) for a neutral
population.

It is usual to take a pre-existing `wild-type' allele (we'll take this
to be $B$) to have fitness $w_B=1$, and the `mutant' ($A$) to have
fitness $w_A=1+s$.  Then, the mean change in the number of mutants in
one generation, given that there are $n(t)$ mutants in generation $t$,
is
\begin{equation}
\frac{\langle n(t+1)\rangle - n(t)}{N} = s \frac{n(t)}{N} \left( 1 -
\frac{n(t)}{N} \right) + O(s^2)\;.
\end{equation}
Denoting the frequency of mutant $A$ alleles as $x$, the Fokker-Planck
equation can be shown (e.g., via a Kramers-Moyal expansion or a
large-$N$ expansion) to be
\begin{equation}
\label{FPEs}
\frac{\partial}{\partial t} P(x,t) = - s \frac{\partial}{\partial x}
x(1-x) P(x,t) + \frac{1}{2N} \frac{\partial^2}{\partial x^2} x(1-x)
P(x,t)
\end{equation}
when the relative fitness of the mutant $s$ is small.  In this
instance, small means $s \ll 1$, and not relative to $1/N$.  Hence, if
at some fixed $s$ one has the (effective) population size $N \gg 1/s$,
the effects of drift can typically be neglected.  Then the mean allele
frequency satisfies the deterministic equation
\begin{equation}
\label{logistic}
\frac{\rmd x}{\rmd t} = s x(1-x)
\end{equation}
and one finds a logistic growth in the number of mutants
\begin{equation}
x(t) = \frac{x(0)}{x(0) + [1 - x(0)] \rme^{-st}} \;.
\end{equation}

Typically it is assumed that the stochastic effects of drift are
important only when one of the allele frequencies is small, and so the
logistic growth is taken to be representative of the change in
frequency of a beneficial mutation in a randomly mating population
once its frequency has reached some threshold (see e.g.,
\cite{Durrett04}).  This period of rapid growth is often referred to as
a \emph{selective sweep}.  Of course, a mutant allele is present only
in small numbers when it first appears, and here one can work out the
probability that a mutant allele will be lost due to genetic drift.
To do this one needs to solve the stationary \emph{backward}
Fokker-Planck equation corresponding to (\ref{FPEs})
\begin{equation}
\label{BFPEs}
0 = x(1-x) \left[ s \frac{\rmd}{\rmd x} Q(x) + \frac{1}{2N}
  \frac{\rmd^2}{\rmd x^2} Q(x) \right]
\end{equation}
subject to the boundary conditions $Q(0)=0$ and $Q(1)=1$, since $Q(x)$
is the probability that a mutant allele becomes fixed given that its
initial frequency is $x$.  Integrating once gives $Q'(x) \propto
\rme^{-2Nsx}$ and again gives $Q(x) = A \rme^{-2Nsx} + B$ where $A$
and $B$ are to be fixed by the boundary conditions.  This yields
\cite{Kimura62} 
\begin{equation}
\label{Qs}
Q(x) = \frac{1 - \rme^{-2Nsx}}{1 - \rme^{-2Ns}} \;.
\end{equation}
In particular, if there is initially only a single mutant, $x=1/N$, in
the limit of an infinite population one has
\begin{equation}
\lim_{N\to\infty} Q(1/N) = \left\{ \begin{array}{ll} 2s &
  0 < s \ll 1 \\ 0 & s \le 0 \end{array} \right.
\end{equation}
where we recall that to obtain (\ref{FPEs}) it was assumed that $s$ is
small.  From the backward Fokker-Planck equation (\ref{BFPEs}) one can
also find the mean number of generations $\tau$ until fixation of a
selectively advantageous allele. It is given approximately as
\begin{equation}
\label{taus}
\tau = \frac{2 \ln N}{s}
\end{equation}
when the combination $Ns$ is large (e.g., $s$ some fixed value and
$N\to\infty$) \cite{Ewens04}.

One may ask how selection affects the backward-time formulation of the
population dynamics that is couched in terms of genealogies.  It turns
out that the complexity of the calculations increases considerably,
because as one goes back in time one needs to keep track of the number
of alleles of each type present in the population.  One can sometimes
find situations in which there is an equilibrium in these frequencies,
for example, when a selectively disadvantageous allele ($s<0$) is
maintained in a population due to recurrent mutations generating it
\cite{Kaplan88}.  When there is a selective sweep, it can be shown
that the relevant genealogies are those that have \emph{multiple}
lineages coalescing simultaneously \cite{Durrett05}---compare with
the case of neutral evolution when the probability of a triple merger
is suppressed by a factor of $1/N$ compared to that of a pairwise
coalescence.  This is consistent with the fact that for fixation to be
achieved on a timescale sublinear in $N$ (\ref{taus}): recall, that in
a state of fixation \emph{all} $N$ individuals share a common
ancestor.  One way to formulate the genealogical process with
selection is through the ``$\Lambda$-coalescent'' \cite{Pitman99};
meanwhile, certain aspects have recently been elucidated by
considering the properties of noisy travelling waves \cite{Brunet06}.

Given the discussion of Section~\ref{nonideal}, one may also be
interested in determining how effects of selection and population
subdivision combine.  In a number of ways the situation is rather
similar to the neutral case, at least if the selective advantage $s$
(or disadvantage, if negative) is not spatially dependent.  For
example, when migration is strong one expects fixation probabilities
and times to be given (\ref{Qs}) and (\ref{taus}) but with $N$ being
an effective population size of the order of the total population size
\cite{Slatkin81}.  In the slow migration limit, the mean time for a
lineage to hop between subpopulations $\sim N$ is much longer than the
duration of a selective sweep $\sim \ln N$ and so typically each
subpopulation is taken to be fixed in either the wild-type or mutant
state.  One thus calculates fixation probabilities and times by having
wild-type subpopulations invaded by their mutant neighbours (and vice
versa) on the migration timescale, and a flip from the wild-type to
mutant state occurring with the probability given by $Q(1/N)$, and in
the other direction with the same expression but with $s$ replaced by
$-s$.  When conservative migration is in force, it can be shown that
the probability a mutant allele fixes is independent of the location
of the initial mutation \cite{Maruyama70a,Slatkin81}: a similar
situation occurred in the neutral case (although the fixation
probability is different).  To see deviations from this behaviour, one
needs either to introduce additional processes (such as extinction and
recolonisation of subpopulations \cite{Barton93}) or relax the
assumption of conservative migration \cite{Lieberman05}.  In the
latter case one finds that the network structure connecting the
subpopulations strongly influences whether selection or drift is the
dominant process.  For example, star-like structures amplify the
effects of selection and can be constructed such that an advantageous
mutation appearing almost anywhere is guaranteed to fix.  On the other
hand, it is also possible to contrive networks in which fixation
occurs whenever the mutation occurs within a particular subpopulation,
and never if it appears elsewhere.  This latter type of behaviour does
not, in fact, depend on the mutation having a selective advantage:
even neutral mutations appearing in ``well-connected'' parts of a
system can fix with high probability when migration is a
non-conservative process \cite{Blythe07}.

Shifting focus from networks to continuous space, one can write down
an equation for the advance of an advantageous mutation.  Recall that
in the absence of drift, one has the deterministic equation
(\ref{logistic}) for the mutant gene frequency $x$ in a subpopulation.
With isotropic migration in one dimension with a coordinate $u$ one
would augment this equation with a diffusion term
\begin{equation}
\label{FKPP}
\frac{\partial}{\partial t} x(u, t) = D \frac{\partial^2 }{\partial
u^2} x(u, t) + s x(u)[1-x(u)] \;.
\end{equation}
This is known as the Fisher or KPP (Kolmogorov-Petrovksy-Piskunov)
equation \cite{Fisher37,Kolmogorov37} and admits travelling wave 
solutions of the form $x(u,t)= f(u - vt)$ where $v$ is the wave velocity.  
If one anticipates that the leading edge of the wave has an exponential 
decay $f(\xi) = \rme^{-\lambda \xi}$, one finds a range of velocities 
$v = D \lambda + s/\lambda \ge 2 \sqrt{s/D}$ are possible.  It turns out 
that in such an equation, if the interface between the stable and unstable 
phases (here, regions of high and low mutant frequencies) is sufficiently
sharp, the front velocity selected is the smallest allowed
\cite{vanSaarloos03}.  If genetic drift is reintroduced, e.g., by
adding a white noise to (\ref{FKPP}) with zero mean and variance
$x(1-x)/N$ (cf.\ (\ref{FPEs}), but note the usual problems that arise
from introducing multiplicative noise in such an \textit{ad-hoc} way),
one expects strong fluctuations in the leading edge of the travelling
wave as a consequence of the possibility that a newly introduced
mutant may go extinct.  This causes a very small shift in the velocity
of the front and a more diffuse profile than in the deterministic case
\cite{Brunet01}.

\begin{figure}
\begin{center}
\includegraphics[scale=0.33]{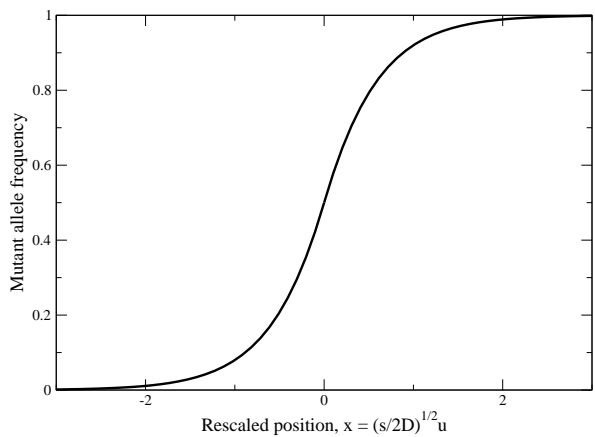}
\end{center}
\caption{\label{cline} Variation of mutant allele frequency with
  position in a steady state where selection is balanced by
  migration.  The mutant allele has a selective advantage $s$ at
  positions $u>0$, a disadvantage $-s$ at positions $u<0$.  Migration
  is characterised by a diffusion constant $D$.  Although hard to
  discern on the plot, there is a discontinuity in the gradient of the
  allele frequencies at $u=0$.}
\end{figure}

One can also consider models with spatially varying fitnesses.  A
simple example would be if a mutant had fitness advantage $+s$ at
position $u>0$ and disadvantage $-s$ at $u<0$ ($s$ here is taken to be
a positive number) \cite{Haldane48}.  At $u\to\infty$ one anticipates
that the mutant allele would be fixed ($x=1$), whilst at $u\to-\infty$
only the wild type would be found ($x=0$).  Using (\ref{FKPP}) the
steady state can be found by solving the nonlinear equation
\begin{equation}
D \frac{\rmd^2}{\rmd u^2} x(u) = \pm s x(u) [ 1 - x(u) ]
\end{equation}
where the positive sign is taken for $u<0$ and the negative sign for
$u>0$.  The symmetry of the problem implies that $x(u) = 1 - x(-u)$,
and in particular that $x(0)=\frac{1}{2}$.  However, one finds the
step in the fitness landscape at $u=0$ induces a discontinuity in the
gradient of $x(u)$ at that point.  To see this, one must solve this
differential equation which is achieved by multiplying both sides by
$\frac{\rmd x}{\rmd u}$ and integrating twice.  This procedure leads
to \cite{Haldane48}
\begin{equation}
x(u) = \left\{ \begin{array}{ll}
\frac{3}{2} \left[ 1 - \tanh^2 \left( \frac{1}{2} \sqrt{\frac{s}{D}}
  u - \kappa \right) \right] & u<0 \\
\frac{3}{2} \tanh^2 \left( \frac{1}{2} \sqrt{\frac{s}{D}}
  u + \kappa \right) - \frac{1}{2} & u>0
\end{array}
\right.
\end{equation}
where
\begin{equation}
\kappa = \frac{1}{2} \ln \left( \frac{ \sqrt{3} + \sqrt{2} }{ \sqrt{3} -
  \sqrt{2} } \right) = 1.146216\ldots\;.
\end{equation}
This function is plotted in Fig.~\ref{cline}.  The spatial variation
of allele frequencies due to a balance between migration and a
changing fitness landscape has been called a \emph{cline} \cite{Huxley38}.

Finally one can consider variation in fitness not in real space, but
in the space of genotypes (i.e., possible allele combinations).  Let
us return to a randomly-mating diploid population, in which a fraction
$x_i$ of all genes are of allele $A_i$, $i=1,2,\ldots,n$.  It is
fairly straightforward to show \cite{Crow70} that, in the absence of drift, 
the change in allele frequency per generation is
\begin{equation}
\Delta x_i = \frac{x_i(1-x_i)}{2} \frac{\partial
}{\partial x_i} \ln \bar{w}
\end{equation}
where the mean fitness of the population is
\begin{equation}
\bar{w} = \sum_{i,j} x_i x_j w_{ij}
\end{equation}
and $w_{ij}$ is the fitness of an individual carrying the pair of
alleles $(A_i, A_j)$.  One can therefore view the function $f(x_1,x_2,
\ldots, x_n) = -\ln \bar{w}$ as a kind of free energy defined over the
space of allele frequencies which the evolutionary dynamics seeks to
minimise.  If we now extend the discussion to multiple gene
\emph{loci} (i.e., genes coding for different traits) with
interactions between them, it is possible for an extremely rugged
fitness landscape in the space of all possible allele frequencies to
emerge \cite{Whitlock95}.  One can then find a transition between two
distinct regimes induced by a change in mutation rate: if the total
number of mutants appearing per generation across the whole population
(which is governed by the population size $N$ multiplied by the
mutation rate $u$) is small, fixation of a selectively advantageous
allele is likely to occur before the onset of the next mutation (see
e.g, \cite{Nowak82,Jain07}).  Thus, all individuals in the population
are then likely to have the same genotype, and one which tends towards
a local fitness maximum.  On the other hand, if the number of mutants
appearing per generation is large, and one is likely to see many
different genotypes in the population, each corresponding to a
different fitness maximum.  In the latter case the individuals are
said to form a \emph{quasi-species} \cite{Eigen77,Nowak82}

\section{Conclusion}
\label{conclusion}

This article has been a review of the ideas and formalism used to
model stochastic processes in fields that statistical physicists are
not typically acquainted with, specifically population genetics,
ecology and linguistics.  As a consequence, some parts of the
discussion will seem familiar, other parts will not.  We have tried,
and we hope that we have succeeded, to explain the background ideas
and motivation, since this will be the greatest obstacle to
understanding among a readership of statistical physicists. On the
other hand the degree of mathematical sophistication that has been
assumed is greater than would be typical outside physics or
mathematical biology.  This makes review quite different to others in
the same area, and while we expect the readership to be mainly
statistical physicists, we hope that some of those working
particularly in ecology and linguistics will find our approach to
their subject interesting and stimulating.
 
In our discussions of the mathematical models, we have mostly used the
language of population genetics, but through of the mappings discussed
in Section~\ref{PG_E_L} the results obtained are more widely relevant.
As the evolutionary paradigm becomes even more widely applied, there
may be other areas in which analogies can be drawn.  It is interesting
how neutral processes turn out to have greater importance in all three
areas we discussed; at the very least neutral theories can be thought
of a null models, against which data and other models can be compared.
Most textbooks in population genetics begin their discussion of
genetic drift with the Wright-Fisher model, although for physicists
the use of non-overlapping generations and a ``time'' measured in
number of generations will not appear so natural.  The Moran model,
which has exactly the same limit when the number of genes become
large, is far more familiar, resembling a birth/death processes where
a death is immediately followed by a birth.  In addition, the
continuous time limit may easily be taken, leading to a master
equation of a kind well-known in statistical physics.

We spent some time explaining the relationship between the
discrete-time approach, based on transition probabilities, and the
continuous-time master equation approach, based on processes occurring
independently at a given rate.  The use of master equations is
apparently rare in the population genetics literature, which is one of
the reasons why we have gone into this approach in such detail.
Furthermore, this latter formalism usually turns out to be more
efficient.  For example, in his book Hubbell~\cite{Hubbell01}
formulated his model by analogy with the corresponding discrete-time
genetic models, and had to perform simulations to generate the
stationary probability distribution.  However, if the model is
formulated as a master equation, the stationary probability
distribution can be obtained analytically~\cite{McKane04}.  However,
the relationship between the discrete- and continuous-time
formulations is not necessarily straightforward.  For example, in
Section~\ref{continuous_time} we outlined two ways in which mutation
can be included in a continuous-time formulation.  One is to have
combined birth, death and mutation events, with on average one
occurring per unit time.  The other is to have mutation as a
spontaneous Poisson process (like radioactive decay).  We showed that
whilst all discrete-time (Moran) processes with mutation could be
realised using the former approach, restrictions were imposed on the
latter.

Another aspect of the article, which to our knowledge has not been
elucidated elsewhere in the context of these problems, is the emphasis
we have put on starting with individual (or gene or token) based
models and deriving the corresponding mesoscopic model, where the
limit of the number of individuals, $N$, tends to infinity.
Simulations are frequently carried out using individuals, and
moreover, as is well-known, an infinity of individual based models
will give the same mesoscopic model in the large $N$ limit.  For
example, the language evolution model which we have alluded to, was
first discussed in a mesoscopic context~\cite{Baxter06}, but this
analysis was restricted to quite simple situations, for instance where
all speakers spoke equally with each other.  When more complex
situations are analysed, such as allowing for a non-trivial network
topology for links between speakers~\cite{Blythe07}, the
individual-based picture becomes far more useful.  In particular, the
backwards-time or coalescent approach is a far more efficient way of
performing simulations. In some cases, such as understanding trends in
how fixation time changes with network structure, which require
simulations of quite large networks, the coalescent formulation is
indispensable.

In genetics, ecology or language, just as in physics, reality cannot
be described by ideal models; there will be a multitude of ways in
which real systems deviate from the ideal models created by scientists
when they first enter a field.  One of the methods that has been
devised by population geneticists to deal with this will be very
familiar to physicists.  This is to characterise a non-ideal system by
a few parameters, which will hopefully, if chosen correctly, capture
the essence of the system. It may be that a simple model can then be
utilised, but with these parameters built in.  An example is the
effective population size, $N_e$, discussed in Section
\ref{effective_pop}, which reflects how the non-ideal nature of the
system changes the effective value of $N$: the effective size of the
real population being the number of individuals in the ideal
population which gives the same magnitude for the quantity of
interest. The effective population size and inbreeding coefficient,
discussed in Section \ref{inbreeding_coeff}, are extremely widely
used, but unfortunately not in a very systematic way.  We have
attempted to illustrate their use, and to draw attention to some of
the confusion which exists in the literature, but it is still the case
that those using these useful concepts will need to be vigilant when
reading the literature on them.

This last observation opens up some questions that may be worth of
further study.  First one can ask whether inbreeding coefficients and
the effective population size really adequately characterise the
evolution of a non-ideal population.  Are there other approximation
schemes that might allow deviation from ideal behaviour to be
described in a systematic and controlled way?  Secondly, the
backward-time description of neutral evolution will be recognised by
nonequilibrium statistical physicists as the reaction-diffusion
process $A+A\to A$.  Although exact solutions for this process exist
for certain mathematically convenient geometries (such as continuous
real space \cite{Peliti86} and, as we have seen, the complete graph) there
is interest in extending the analysis to heterogeneous graphs, which
thus far have been treated only using approximate methods
\cite{Sood05,Antal06,Blythe07}.  Such solutions would be useful as
they would obtain more detail about the evolution of the system, for
example, the manner in which an innovation propagates through the
system \textit{en route} to fixation. There are many other questions 
which would benefit from a systematic study of the type that statistical
physicists are well-placed to carry out. We hope that this article has
stimulated some readers to find out more about these fascinating applications 
of stochastic methods and to contribute to the development of the field.

\section*{Acknowledgements}

We wish to thank David Alonso and Alan Bray for useful discussions.
RAB holds a Personal Research Fellowship awarded by the Royal Society
of Edinburgh.

\renewcommand{\theequation}{\Alph{section}.\arabic{equation}}
\setcounter{section}{0} \setcounter{equation}{0}

\begin{appendix}

\section{The large $N$ expansion of the master equation for $M$ alleles}
\label{App_A}

In this Appendix we give details of carrying out the large $N$ expansion 
beginning from the master equations involving $M>2$ alleles, considered 
in Sections \ref{M_alleles} and \ref{L_and_M}.

We begin by expanding out the terms containing the transitions rates
(\ref{rates_M_1_2})---which are those involving the type $M$ 
allele---letting $x_{\alpha} = n_{\alpha}/N$, and expanding all other 
terms in powers of $1/N$. This is the procedure used to obtain 
Eq.~(\ref{M_2_largeN1}), and the analogous expression in this case
\begin{displaymath}
= \sum^{M-1}_{\alpha=1} \left[ \left( x_{\alpha} - \frac{1}{N} \right) 
\left( 1 - \sum^{M-1}_{\beta=1} x_{\beta} + \frac{1}{N} \right) 
\left\{ P - \frac{1}{N} \frac{\partial P}{\partial x_{\alpha}} + 
\frac{1}{2N^2} \frac{\partial^{2} P}{\partial x^{2}_{\alpha}} \right\} \right. 
\end{displaymath}
\begin{displaymath}
\left. - x_{\alpha} \left( 1 - \sum^{M-1}_{\beta=1} x_{\beta} \right) P \right]
\end{displaymath}
\begin{displaymath}
+ \sum^{M-1}_{\alpha=1} \left[ \left( x_{\alpha} + \frac{1}{N} \right) 
\left( 1 - \sum^{M-1}_{\beta=1} x_{\beta} - \frac{1}{N} \right) 
\left\{ P + \frac{1}{N} \frac{\partial P}{\partial x_{\alpha}} + 
\frac{1}{2N^2} \frac{\partial^{2} P}{\partial x^{2}_{\alpha}} \right\} \right.
\end{displaymath}
\begin{displaymath}
\left. - x_{\alpha} \left( 1 - \sum^{M-1}_{\beta=1} x_{\beta} \right) P \right]
= \frac{1}{N^2} \sum^{M-1}_{\alpha=1} \left[ x_{\alpha} \left( 1 - 
\sum^{M-1}_{\beta=1} x_{\beta} \right) \frac{\partial^{2} P}
{\partial x^{2}_{\alpha}} \right. 
\end{displaymath}
\begin{equation}
\left. - 2 x_{\alpha} \frac{\partial P}{\partial x_\alpha} 
+ 2 \left( 1 - \sum^{M-1}_{\beta=1} x_{\beta} \right) 
\frac{\partial P}{\partial x_{\alpha}} - 2 P \right]\,,
\label{first_M_1}
\end{equation}
neglecting terms of order $1/N^3$ and higher. Carrying out the same procedure 
for the terms containing the transition rates (\ref{rates_M_1_1}) gives an 
expression
\begin{eqnarray}
&=& \sum^{M-1}_{\alpha=1} \sum^{M-1}_{\beta \neq \alpha} 
\left[ \left( x_{\alpha} - \frac{1}{N} \right)\left( x_{\beta} + 
\frac{1}{N} \right) \left\{ P - \frac{1}{N} \frac{\partial P}
{\partial x_{\alpha}} + \frac{1}{N} \frac{\partial P}
{\partial x_{\beta}} \right. \right. \nonumber \\
&+& \left. \left. \frac{1}{2N^2} \frac{\partial^{2} P}{\partial x^{2}_{\alpha}}
+ \frac{1}{2N^2} \frac{\partial^{2} P}{\partial x^{2}_{\beta}} 
- \frac{1}{N^2} \frac{\partial^{2} P}{\partial x_{\alpha} 
\partial x_\beta} \right\} - x_{\alpha} x_{\beta} P \right] \nonumber \\
&=& \frac{1}{N^2} \sum_{\alpha=1} \sum^{M-1}_{\beta \neq \alpha} 
\left[ x_{\alpha} x_{\beta} \frac{\partial^{2} P}{\partial x_{\alpha}^{2}} 
- x_{\alpha} x_{\beta} \frac{\partial^{2} P}{\partial x_{\alpha} 
\partial x_{\beta}} + 2 x_{\alpha} \frac{\partial P}{\partial x_{\beta}} 
- 2 x_{\alpha} \frac{\partial P}{\partial x_{\alpha}} - P \right]
\label{second_M_1}
\end{eqnarray}
again neglecting terms of order $1/N^3$ and higher. Adding 
Eqs.~(\ref{first_M_1}) and (\ref{second_M_1}) gives for the master equation
\begin{eqnarray}
\frac{\partial P}{\partial t} &=& \frac{1}{N^{2}} \left[ -M(M-1) P
+ 2 \sum^{M-1}_{\alpha=1} \left\{ \left( 1 - M x_{\alpha} \right) 
\frac{\partial P}{\partial x_\alpha} \right\} \right. \nonumber \\
&+& \left. \sum^{M-1}_{\alpha=1} x_{\alpha} \left( 1 - x_{\alpha} \right)  
\frac{\partial^{2} P}{\partial x_{\alpha}^{2}} - \sum^{M-1}_{\alpha=1}
\sum^{M-1}_{\beta \neq \alpha} x_{\alpha} x_{\beta} \frac{\partial^{2} P}
{\partial x_{\alpha} \partial x_{\beta}} \right]
+ {\cal O} \left( \frac{1}{N^3} \right) \nonumber \\
&=& \frac{1}{N^2} \left\{ \sum^{M-1}_{\alpha=1} \frac{\partial^{2} }
{\partial x_{\alpha}^{2}} \left[ x_{\alpha} \left( 1 - x_{\alpha} \right) 
P \right] - \sum^{M-1}_{\alpha=1} \sum^{M-1}_{\beta \neq \alpha} 
\frac{\partial^{2} }{\partial x_{\alpha} \partial x_{\beta}} 
\left[ x_{\alpha} x_{\beta} P \right] \right\}\,,
\label{third_M_1} 
\end{eqnarray}
up to terms of order $1/N^3$ and higher. As in previous cases, if we define 
a new time $\tau = 2t/N^2$, and take $N \to \infty$, then we find the 
result Eq.~(\ref{diffusion_M_1}).

In the main text we discussed two different ways of including mutations. In
the first type of model, mutation rates were given by Eq.~(\ref{muta}). This
equation may be written as
\begin{equation}
T(n_1 \ldots n_{\alpha}+1 \ldots n_{\beta}-1 \ldots n_{M-1}|
\underline{n}) = \left( 1 - \rho_{\alpha} \right) \frac{n_\alpha}{N}
\frac{n_\beta}{N} + \sigma_{\alpha} (\underline{n}/N) \frac{n_{\beta}}{N}\,,
\label{muta_app}
\end{equation}
where
\begin{equation}
\rho_{\alpha} = \sum_{\gamma \neq \alpha} u_{\gamma \alpha} \ \ {\rm and}
\ \ \sigma{_\alpha} (\underline{x}) = \sum_{\delta \neq \alpha} 
u_{\alpha \delta} x_{\delta}\,.
\label{r_and_s}
\end{equation}
Suppose that we first look at the case where $\alpha, \beta=1,\ldots,M-1$ in 
the transition rate (\ref{muta_app}). This gives the following contribution to 
the right-hand side of the Fokker-Planck equation:
\begin{displaymath}
\sum^{M-1}_{\alpha=1} \sum^{M-1}_{\beta \neq \alpha} 
\left[ \left( 1 - \rho_{\alpha} \right) \left( x_{\alpha} - 
\frac{1}{N} \right) + \sigma_{\alpha} (\underline{x}) \right] \left[ x_{\beta}
+ \frac{1}{N} \right] \left( P + \frac{1}{N} \frac{\partial P}
{\partial x_{\beta}} - \frac{1}{N} \frac{\partial P}
{\partial x_{\alpha}} \right) 
\end{displaymath}
\begin{displaymath}
- \sum^{M-1}_{\alpha=1} \sum^{M-1}_{\beta \neq \alpha} 
\left[ \left( 1 - \rho_{\alpha} \right) x_{\alpha} + 
\sigma_{\alpha} (\underline{x}) \right] x_{\beta} P
+ {\cal O} \left( \frac{1}{N^2} \right)\,. 
\end{displaymath}
The terms involving only the mutation rates are
\begin{displaymath}
\frac{1}{N} \sum^{M-1}_{\alpha=1} \sum^{M-1}_{\beta \neq \alpha} 
\left[ - \rho_{\alpha} x_{\alpha} + \rho_{\alpha} x_{\beta} 
+ \sigma_{\alpha} (\underline{x}) \right] P 
\end{displaymath}
\begin{equation}
+ \frac{1}{N} \sum^{M-1}_{\alpha=1} \sum^{M-1}_{\beta \neq \alpha} 
\left[ - \rho_{\alpha} x_{\alpha} x_{\beta} + \sigma_{\alpha} (\underline{x})
x_{\beta} \right] \left\{ \frac{\partial P}{\partial x_{\beta}} 
- \frac{\partial P}{\partial x_{\alpha}} \right\} + {\cal O} 
\left( \frac{1}{N^2} \right). 
\label{first_M_1_Mut}
\end{equation}
The analogous expression when $\alpha = 1,\ldots,M-1$ and $\beta=M$ is:
\begin{displaymath}
\sum^{M-1}_{\alpha=1} \left[ \left( 1 - \rho_{\alpha} \right) \left( x_{\alpha}
- \frac{1}{N} \right) + \sigma_{\alpha} (\underline{x}) \right] 
\left[ 1 - \sum^{M-1}_{\gamma =1} x_{\gamma} + \frac{1}{N} \right] 
\left( P - \frac{1}{N} \frac{\partial P}{\partial x_{\alpha}} \right) 
\end{displaymath}
\begin{displaymath}
- \sum^{M-1}_{\alpha=1} \left[ \left( 1 - \rho_{\alpha} \right) x_{\alpha} + 
\sigma_{\alpha} (\underline{x}) \right] \left[ 1 - \sum^{M-1}_{\gamma=1} 
x_{\gamma} \right] P + {\cal O} \left( \frac{1}{N^2} \right)\,, 
\end{displaymath}
and when $\beta = 1,\ldots,M-1$ and $\alpha=M$ is:
\begin{displaymath}
\sum^{M-1}_{\beta=1} \left[ \left( 1 - \rho_{M} \right) \left( 1 -
\sum^{M-1}_{\gamma=1} x_{\gamma} - \frac{1}{N} \right) + 
\sigma_{M} (\underline{x}) \right] \left[ x_{\beta} + \frac{1}{N} \right] 
\left( P + \frac{1}{N} \frac{\partial P}{\partial x_{\beta}} \right) 
\end{displaymath}
\begin{displaymath}
- \sum^{M-1}_{\beta=1} \left\{ \left( 1 - \rho_{M} \right) \left( 1 -
\sum^{M-1}_{\gamma=1} x_{\gamma} \right) + \sigma_{M} 
(\underline{x}) \right\} x_{\beta} P + {\cal O} \left( \frac{1}{N^2} \right)\,,
\end{displaymath}
which lead to the analogous expressions to Eq.~(\ref{first_M_1_Mut}):
\begin{displaymath}
\frac{1}{N} \sum^{M-1}_{\alpha=1} \left[ - \rho_{\alpha} x_{\alpha} + 
\rho_{\alpha} \left( 1 - \sum^{M-1}_{\gamma=1} x_{\gamma} \right) + 
\sigma_{\alpha} (\underline{x}) \right] P 
\end{displaymath}
\begin{equation}
- \frac{1}{N} \sum^{M-1}_{\alpha=1} \left[ - \rho_{\alpha} x_{\alpha} 
\left( 1 - \sum^{M-1}_{\gamma=1} x_{\gamma} \right) + \sigma_{\alpha} 
(\underline{x}) \left( 1 - \sum^{M-1}_{\gamma=1} x_{\gamma} \right) \right] 
\frac{\partial P}{\partial x_{\alpha}} + {\cal O} 
\left( \frac{1}{N^2} \right). 
\label{second_M_1_Mut}
\end{equation}
and
\begin{displaymath}
\frac{1}{N} \sum^{M-1}_{\beta=1} \left[ \rho_{M} x_{\beta} - 
\rho_{M} \left( 1 - \sum^{M-1}_{\gamma=1} x_{\gamma} \right) + 
\sigma_{M} (\underline{x}) \right] P 
\end{displaymath}
\begin{equation}
+ \frac{1}{N} \sum^{M-1}_{\beta=1} \left[ - \rho_{M} \left( 1 - 
\sum^{M-1}_{\gamma=1} x_{\gamma} \right) + 
\sigma_{\alpha} (\underline{x}) \right] x_{\beta} \frac{\partial P}
{\partial x_{\beta}} + {\cal O} \left( \frac{1}{N^2} \right)\,. 
\label{third_M_1_Mut}
\end{equation}
From the expressions (\ref{first_M_1_Mut})-(\ref{third_M_1_Mut}), the term 
involving mutations in the Fokker-Planck equation can be found. For 
example, let us focus on the mutation rates involving $A_\alpha$ and $A_\beta$
with $\alpha, \beta=1,\ldots,M-1$ and $\alpha \neq \beta$. After some 
straightforward algebra, the terms from Eqs.~(\ref{first_M_1_Mut}) and
(\ref{second_M_1_Mut}) which involve $P$, but not derivatives of $P$ are
\begin{equation}
\frac{1}{N} \sum^{M-1}_{\alpha=1} \left[ \rho_{\alpha} \left( 1 - 
M x_{\alpha} \right) + M \sigma_{\alpha} (\underline{x}) \right] P 
= \frac{1}{N} \sum^{M-1}_{\alpha=1} \sum_{\gamma \neq \alpha} u_{\gamma \alpha}
P\,,
\label{fourth_M_1_Mut}
\end{equation}
using Eq.~(\ref{r_and_s}). The terms from Eqs.~(\ref{first_M_1_Mut}) and
(\ref{second_M_1_Mut}) which involve derivatives of $P$ are more complicated,
but eventually yield
\begin{displaymath}
\frac{1}{N} \sum^{M-1}_{\alpha=1} \left[ \rho_{\alpha} x_{\alpha} - 
\sigma_{\alpha} (\underline{x}) \right] \frac{\partial P}{\partial x_\alpha}
\end{displaymath}
\begin{equation}
= \frac{1}{N} \sum^{M-1}_{\alpha=1} \left[ \sum_{\gamma \neq \alpha} 
u_{\gamma \alpha} x_{\alpha} - \sum_{\delta \neq \alpha} u_{\alpha \delta} 
x_{\delta} \right] \frac{\partial P}{\partial x_\alpha}\,.
\label{fifth_M_1_Mut}
\end{equation}
We may combine Eqs.~(\ref{fourth_M_1_Mut}) and (\ref{fifth_M_1_Mut}) to give
\begin{equation}
- \frac{1}{N} \sum^{M-1}_{\alpha=1} \frac{\partial }{\partial x_\alpha}
\left[ \sum_{\delta \neq \alpha} u_{\alpha \delta} x_{\delta} - 
\sum_{\gamma \neq \alpha} u_{\gamma \alpha} x_{\alpha} \right] P + {\cal O} 
\left( \frac{1}{N^2} \right)\,.
\label{sixth_M_1_Mut}
\end{equation}
Introducing the scaled mutation rates defined in 
Eq.~(\ref{scaled_rates_general}), this expression equals
\begin{equation}
- \frac{2}{N^2} \sum^{M-1}_{\alpha=1} \frac{\partial }{\partial x_\alpha}
\left[ \sum_{\delta \neq \alpha} {\cal U}_{\alpha \delta} x_{\delta} - 
\sum_{\gamma \neq \alpha} {\cal U}_{\gamma \alpha} x_{\alpha} \right] P 
+ {\cal O} \left( \frac{1}{N^3} \right)\,.
\label{seventh_M_1_Mut}
\end{equation}
Using the rescaled time $\tau = 2t/N^{2}$ and letting $N \to \infty$,
we see that we obtain the mutation terms displayed in 
Eqs.~(\ref{diffusion_M_Mut_1}) and Eqs.~(\ref{A_M_1}), at least in the
sector where $\alpha, \beta=1,\ldots,M-1$. In a similar way, the mutation 
terms where the allele $A_M$ is involved may be derived from 
Eqs.~(\ref{second_M_1_Mut}) and (\ref{third_M_1_Mut}). One again finds the 
result given by Eqs.~(\ref{diffusion_M_Mut_1}) and Eqs.~(\ref{A_M_1}).

The form of the mutation terms in the Fokker-Planck equation in the second 
scheme we investigated, defined by the transition rates (\ref{mut_second}),
follows from the results obtained above for the first scheme. This is 
because if one takes $u_{\alpha \beta} = u_\alpha$ for all 
$\beta \neq \alpha$, then the transition rates of the two approaches are 
identical, although in the second case there is the restriction that 
$\sum^{M}_{\delta=1} u_{\delta} < 1$. We therefore find the same Fokker-Planck 
equation (\ref{diffusion_M_Mut_1}), but now with Eq.~(\ref{A_M_1}) taking
the form 
\begin{equation}
{\cal A}_{\alpha} (\underline{x}) = \sum_{\beta=1}^{M} \left({\cal
U}_{\alpha} x_{\beta} - {\cal U}_{\beta} x_{\alpha}\right) =
{\cal U}_{\alpha} - \sum^{M}_{\beta=1} {\cal U}_{\beta} x_{\alpha}\,,
\end{equation}
which is Eq.~(\ref{cal_A_mod}). When written out in terms of the scaled 
mutation rates, the condition on the $u_\delta$ in the second case reads
\begin{equation}
\sum^{M}_{\delta=1} {\cal U} < \frac{N}{2}\,,
\end{equation}
which, since the rescaled rates are numerically small and we have taken the 
limit $N \to \infty$, is always satisfied.

Finally, we consider the most general case where there are $L$ islands and 
$M$ alleles. As we remarked in the main text, this involves only minor 
modifications of calculations which have been carried out previously. This is
because, since both the mutation and migration rates are scaled by a factor 
of $N$, any product of these rates will give a term of order $1/N^{3}$, and so
will not contribute to the Fokker-Planck equation. Therefore we may consider
mutation and migration separately, and simply need to insert appropriate 
indices on the contributions already calculated.

First let, us consider Eq.~(\ref{M_M_mig_probs_L}), which does not include 
mutation. If we begin by ignoring the migration terms $g_{i j}$ with 
$i \neq j$, then we see that the transition rates will be of the form given
in Eqs.~(\ref{rates_M_1_1}) and (\ref{rates_M_1_2}), but multiplied by $f_i$
and with an index $i$ included. Thus the result of a $1/N$ expansion is as 
in Eq.~(\ref{third_M_1}), but with an index $i$ and $f_i$ included:
\begin{equation}
\frac{f_i}{N^2} \left\{ \sum^{M-1}_{\alpha=1} 
\frac{\partial^{2} }{\partial x_{i \alpha}^{2}} \left[ x_{i \alpha} 
\left( 1 - x_{i \alpha} \right) P \right] - \sum^{M-1}_{\alpha=1}
\sum^{M-1}_{\beta \neq \alpha} \frac{\partial^{2} }{\partial x_{i \alpha} 
\partial x_{i \beta}} \left[ x_{i \alpha} x_{i \beta} P \right] \right\}\,,
\label{first_M_L}
\end{equation}
up to terms of order $1/N^{3}$ and higher. This has to be summed over all $i$.

Returning to Eq.~(\ref{M_M_mig_probs_L}), the transition rates which 
involve migration can be found by analogy with the discussion in Section 
\ref{L_islands}, and in particular Eq.~(\ref{M_mig_rates_L}), with $\alpha$ 
and $\beta$ indices for the allele labels. These appear as in 
Eqs.~(\ref{first_M_1}) and (\ref{second_M_1}). The result is: 
\begin{eqnarray}
G_{i j} \sum^{M-1}_{\alpha=1} \sum^{M-1}_{\beta \neq \alpha} 
\left[ \left( x_{i \beta} + \frac{1}{N} \right) x_{j \alpha} 
\left\{ P - \frac{1}{N} \frac{\partial P}{\partial x_{i \alpha}} + 
\frac{1}{N} \frac{\partial P}{\partial x_{i \beta}} \right\} \right.
\nonumber \\
\left. - x_{i \beta} x_{j \alpha} P \right] + G_{i j} \sum^{M-1}_{\alpha=1} 
\left[ \left( 1 - \sum^{M-1}_{\gamma=1} x_{i \gamma} + \frac{1}{N} \right) 
x_{j \alpha} \left\{ P - \frac{1}{N} \frac{\partial P}
{\partial x_{i \alpha}} \right\} \right. \nonumber \\
\left. - \left( 1 - \sum^{M-1}_{\gamma=1} x_{i \gamma} \right) x_{j \alpha} 
P \right] + G_{i j} \sum^{M-1}_{\beta=1} \left[ \left( x_{i\beta} + 
\frac{1}{N} \right) \left( 1 - \sum^{M-1}_{\gamma=1} x_{j \gamma} \right) 
\right. \nonumber \\
\left. \times \left\{ P + \frac{1}{N} \frac{\partial P} 
{\partial x_{i \beta}} \right\} - x_{i \beta} 
\left( 1 - \sum^{M-1}_{\gamma=1} x_{j \gamma} \right) P \right]
+ {\cal O} \left( \frac{1}{N^2}\right) \nonumber \\
= G_{i j} \frac{1}{N} \left[ \left( M-1 \right) P + 
\sum^{M-1}_{\alpha=1} \left( x_{i \alpha} - x_{j \alpha} \right) 
\frac{\partial P}{\partial x_{i \alpha}} \right] 
+ {\cal O} \left( \frac{1}{N^2}\right) \nonumber \\
= G_{i j} \frac{1}{N}\,\sum^{M-1}_{\alpha=1} \frac{\partial }
{\partial x_{i \alpha}} \left[ \left( x_{i \alpha} - x_{j \alpha} \right) 
P \right] + {\cal O} \left( \frac{1}{N^2}\right)\,.
\label{second_M_L}
\end{eqnarray} 
Scaling the migration rate by introducing ${\cal G}_{i j} = NG_{i j}/2$, adding
an analogous term which represents migration from island $i$ to island $j$,
and summing over all pairs of islands $\langle i j \rangle$ gives 
\begin{equation}
\frac{2}{N^2} \sum_{\langle i j \rangle} \sum^{M-1}_{\alpha=1} 
\left( {\cal G}_{i j} \frac{\partial }{\partial x_{i \alpha}} - 
{\cal G}_{j i}\frac{\partial }{\partial x_{j \alpha}} \right) 
\left[ \left( x_{i \alpha} - x_{j \alpha} \right) P \right]
+ {\cal O} \left( \frac{1}{N^3}\right)\,.
\label{third_M_L}
\end{equation} 
The third and final term is found from Eq.~(\ref{M_M_Mut_mig_probs_L})
ignoring the migration terms $g_{i j}$ with $i \neq j$. This is as in 
Eq.~(\ref{muta}), but multiplied by $f_i$ and with an index $i$ added to the
$n_\beta$ and $n_\delta$. From the discussion leading to 
Eq.~(\ref{seventh_M_1_Mut}) we find that the contribution for the first type 
of mutation is
\begin{equation}
- \frac{2}{N^2} f_{i} \sum^{M-1}_{\alpha=1} \frac{\partial }
{\partial x_{i \alpha}} \left[ {\cal A}_{\alpha} (\underline{x}_{i}) P \right] 
+ {\cal O} \left( \frac{1}{N^3} \right)\,,
\label{fourth_M_L}
\end{equation}
This has to be summed over all $i$. The second type of mutation gives the 
same result, but with ${\cal A}_{\alpha} (\underline{x}_{i})$ defined as in
Eq.~(\ref{cal_A_mod}).

Putting Eqs.~(\ref{first_M_L}), (\ref{third_M_L}) and (\ref{fourth_M_L}) 
together gives equation (\ref{diffusion_M_Mut_L}), a general Fokker-Planck 
equation describing a population of genes, of $M$ possible types, subdivided 
between $L$ different islands.
\end{appendix}

\end{document}